\documentclass{SciPost}

\hypersetup{
    colorlinks,
    linkcolor={red!50!black},
    citecolor={blue!50!black},
    urlcolor={blue!80!black}
}

\usepackage[bitstream-charter]{mathdesign}
\usepackage{subcaption}
\usepackage{booktabs}
\usepackage{multirow}
\usepackage[compat=1.1.0]{tikz-feynman}
\tikzfeynmanset{resonance/.style={thick, dashed, color=red}}
\usepackage{cleveref}
\usepackage{dsfont}
\usepackage{physics}
\usepackage{slashed}
\usepackage{comment}

\DeclareSymbolFont{usualmathcal}{OMS}{cmsy}{m}{n}
\DeclareSymbolFontAlphabet{\mathcal}{usualmathcal}

\fancypagestyle{SPstyle}{
\fancyhf{}
\lhead{\colorbox{scipostblue}{\bf \color{white} ~SciPost Physics }}
\rhead{{\bf \color{scipostdeepblue} ~Submission }}

\fancyfoot[C]{\textbf{\thepage}}
}

\newcommand{\gt}{\Tilde{g}}

\newcommand{\hc}{\mathrm{h.c.}} 
 
\newcommand{\SO}{\mathrm{SO}}
\newcommand{\SU}{\mathrm{SU}}
\newcommand{\U}{\mathrm{U}}
\newcommand{\Sp}{\mathrm{Sp}}

\def\lsim{\mathrel{\rlap{\lower4pt\hbox{\hskip1pt$\sim$}}
    \raise1pt\hbox{$<$}}}         
\def\gsim{\mathrel{\rlap{\lower4pt\hbox{\hskip1pt$\sim$}}
    \raise1pt\hbox{$>$}}}

\newcommand{\GeV}{\ensuremath{\,\text{GeV}}}
\newcommand{\pT}{\ensuremath{p_\text{T}}}
\newcommand{\mZ}{\ensuremath{m_Z}}
\begin{document}

\pagestyle{SPstyle}

\begin{center}{\Large \textbf{\color{scipostdeepblue}{
%%%%%%%%%% TODO: Write your article's title here
Vector boson scattering in composite Higgs models
}}}\end{center}

\begin{center}\textbf{
% Use (full) first name (+ middle name initials) + surname format.
% Separate subsequent authors by a comma, omit comma and use "and" for the last author.
% Mark the corresponding author(s) with a superscript symbol in this order
% \star, \dagger, \ddagger, \circ, \S, \P, \parallel, ...
Rosy Caliri\textsuperscript{$\star$},
Manuel Kunkel\textsuperscript{$\dagger$} and
Werner R.\ Porod\textsuperscript{$\ddagger$}
}\end{center}

\begin{center}
%{\bf 1} Institute of Theoretical Physics and Astrophysics, University of Würzburg, Germany\\
Institute of Theoretical Physics and Astrophysics, University of Würzburg, Germany
%%%%%%%%%% END TODO: AFFILIATIONS
% Provide email address of corresponding author(s)
\\[\baselineskip]
$\star$ \href{mailto:rosy.caliri@uni-wuerzburg.de}{\small rosy.caliri@uni-wuerzburg.de}\,,\quad
$\dagger$ \href{mailto:manuel.kunkel@uni-wuerzburg.de}{\small manuel.kunkel@uni-wuerzburg.de}\,,\\
$\ddagger$ \href{mailto:werner.porod@uni-wuerzburg.de}{\small werner.porod@uni-wuerzburg.de} 

\end{center}

\section*{\color{scipostdeepblue}{Abstract}}
\textbf{\boldmath{%
Vector boson scattering is a powerful tool to test the mechanism of electroweak symmetry breaking. 
We focus on Composite Higgs models with an underlying fermionic description and assume that only one pseudo Nambu-Goldstone boson contributes to electroweak symmetry breaking, which we identify as the Higgs state of the Standard Model.
In these scenarios, LHC data imply lower bounds on the predicted resonances below 2 TeV depending on the scenario.
Using $WZjj$ production as a benchmark, we demonstrate that the tails of various
distributions nevertheless show significant deviations from Standard Model predictions.
This holds in particular for the case of longitudinally polarised vector bosons.
}}

\vspace{\baselineskip}

%%%%%%%%%% BLOCK: Copyright information
% This block will be filled during the proof stage, and finilized just before publication.
% It exists here only as a placeholder, and should not be modified by authors.
%\noindent\textcolor{white!90!black}{%
%\fbox{\parbox{0.975\linewidth}{%
%\textcolor{white!40!black}{\begin{tabular}{lr}%
%  \begin{minipage}{0.6\textwidth}%
%    {\small Copyright attribution to authors. \newline
%    This work is a submission to SciPost Physics. \newline
%    License information to appear upon publication. \newline
%    Publication information to appear upon publication.}
%  \end{minipage} & \begin{minipage}{0.4\textwidth}
%    {\small Received Date \newline Accepted Date \newline Published Date}%
%  \end{minipage}
%\end{tabular}}
%}}
%}
%%%%%%%%%% BLOCK: Copyright information

% For convenience during refereeing we turn on line numbers:
% \linenumbers

\vspace{10pt}
\noindent\rule{\textwidth}{1pt}
\tableofcontents
\noindent\rule{\textwidth}{1pt}
\vspace{10pt}

% Write your article contents here, starting from first \section.
% An example structure is given below.

\section{Introduction}
\label{sec:intro}

The experiments at the Large Hadron Collider (LHC) have firmly established the success
of the Standard Model (SM) of particle physics, confirming it as a remarkably accurate framework for
describing fundamental interactions up to the TeV range. 
The discovery of the Higgs boson \cite{ATLAS:2012yve,CMS:2012qbp}, along with the observation of the electroweak production
of massive vector bosons \cite{ATLAS:2018iui,ATLAS:2019cbr,ATLAS:2019thr,CMS:2019uys,CMS:2021qzz,CMS:2025dbm,ATLAS:2025omi,CMS:2026add} and  tri-boson production \cite{ATLAS:2019dny,CMS:2020hjs} are among the most significant recent achievements. 
However, the SM leaves many fundamental questions unanswered, such as the origin of the cosmological baryon asymmetry, the nature
of dark matter and dark energy, and the origin of neutrino masses and mixing. These open problems make the search for
physics beyond the Standard Model (BSM) more pressing than ever.

Despite intensive efforts, there are no direct signs of new physics at the LHC. 
The ATLAS and CMS experiments have set increasingly stringent exclusion limits, pushing the mass scales of hypothetical
new particles into the multi-TeV range. At the same time, these experiments are turning into precision experiments due to
the increasing wealth of data and new analysis strategies, allowing for the search for subtle deviations from the SM. 
One of the most promising places to search for new physics is within the mechanism of electroweak symmetry breaking (EWSB). 
Besides generating the masses of the $W$ and $Z$ bosons, EWSB  also governs how $W$, $Z$ and the Higgs interact at high energies. Moreover, it ensures the unitarity of their scattering amplitudes.
These effects converge in the study of vector boson scattering (VBS) processes, making them a promising target for testing the EW sector, see e.g.~ref.\ \cite{BuarqueFranzosi:2021wrv} for a recent summary.
The longitudinal modes of massive vector bosons arise directly
from the EWSB mechanism, so their behaviour at high energies provides a unique fingerprint of the precise mechanism \cite{Cornwall:1974km,Gounaris:1986cr,Chanowitz:1985hj}.

In this paper we study VBS processes in the context of Composite Higgs (CH) models, in which the Higgs boson emerges as a pseudo-Nambu-Goldstone boson
(pNGB) due to vacuum misalignment in the breaking of the associated global symmetry \cite{Kaplan:1983fs,Kaplan:1983sm}. 
An interesting class of models 
arises from confining four-dimensional gauge theories (hypercolour) with new fermions, which we call hyperfermions in the following \cite{Ferretti:2013kya,Belyaev:2016ftv}.  
The symmetry breaking pattern is determined by the gauge representation of the hyperfermions \cite{Peskin:1980gc,Preskill:1980mz,Cacciapaglia:2014uja}. 
The resulting condensate breaks a global flavour symmetry $G$ of the hyperfermions to an unbroken subgroup $H$, giving rise to pNGBs that live in the coset $G/H$. 
The phenomenological requirement to realise the Higgs field as a bi-doublet $(\mathbf{2},\mathbf{2})$ of the custodial group 
$\SU(2)_L \times \SU(2)_R \subset H$ fixes the minimal dimension of the corresponding coset.
In the electroweak sector, the following minimal cosets are obtained: $G/H=\SU(5)/\SO(5)$ 
\cite{Ferretti:2014qta,Golterman:2015zwa,Ferretti:2016upr,Agugliaro:2018vsu} for real hyperfermions; 
$\SU(4)/\Sp(4)$ \cite{Barnard:2013zea,Ferretti:2013kya,Ferretti:2014qta,Cacciapaglia:2014uja,Galloway:2016fuo,Agugliaro:2016clv} for pseudo-real ones; 
and $\SU(4)\times \SU(4)/\SU(4)$ \cite{Ma:2015gra,Alanne:2017rrs} for complex ones. 

These models predict an extended pNGB sector and several resonances. Among the latter are electroweak spin-1 resonances~\cite{BuarqueFranzosi:2016ooy,Caliri:2024jdk}. 
The cosets mentioned above are symmetric, implying that the spin-1 resonances belong to two categories: 
(i) states decaying into two pNGBs which will be called vector states and (ii) states decaying into three pNGBs which
will be called axial-vectors. 
One can show that some of these 
states mix with the electroweak vector bosons. In particular, it turns out that all three cosets considered contain one charged 
state mixing substantially with the $W$-boson and two neutral states  mixing substantially with the $Z$-boson \cite{BuarqueFranzosi:2016ooy,Caliri:2024jdk}.  
This is a consequence of the fact that in all cases the custodial group is a subgroup of $H$ by construction. We note for completeness that these states 
correspond essentially to those discussed in \cite{Belyaev:2008yj,Becciolini:2014eba}. Their $s$-channel 
production is constrained by existing LHC data \cite{Caliri:2024jdk}. However, these states can still have masses well below 2~TeV as they do not only decay
into SM fermions but also via various other channels:   
$ V\, h$ ($V=W,Z$), two electroweak vector bosons
as well as two  pNGBs.

VBS has been considered as a probe for strongly interacting theories before, see e.g.~\cite{Foadi:2008xj,BuarqueFranzosi:2017prc,Delgado:2019ucx,Garcia-Garcia:2019oig,BuarqueFranzosi:2023xux} for some recent work and references therein. However, in these papers it has been assumed that the spin-1 states are too heavy
to be produced directly and, thus, the corresponding studies have been carried out in the context of effective field theories in which states are integrated out.
Such an approach is clearly not valid if spin-1 resonances can be produced at the LHC as in the scenarios discussed in ref.~\cite{Caliri:2024jdk}.
We therefore investigate the effect of these states on VBS using a consistent picture in which all relations between the 
modified $hVV$, $VVV$ and $VVVV$ couplings are properly taken into account. In particular, we will demonstrate that there are sizeable effects 
in the tails of lepton distributions taking the $WZ$ final state as a particular example.

The paper is structured as follows:
we summarise in \cref{sec:model} the main features of the models considered.
In \cref{sec:vbs} we present the details of the analysis implemented, and in \cref{sec:results} our numerical results. 
Finally, we draw our conclusions in
\cref{sec:conc}. In addition, we collect further model details in \cref{app:model} and results concerning the
$W^+ W^+$ final state in \cref{app:WW}.

\section{Model aspects in view of vector boson scattering}
\label{sec:model}

Composite Higgs models predict spin-1 resonances that mix with the EW gauge bosons, thus modifying vector boson scattering processes with respect to the SM.
In this section we summarise the relevant model aspects and describe the benchmark parameter points and processes used in the collider study.

\subsection{Relevant states and couplings}

The models proposed in ref.~\cite{Ferretti:2016upr} predict that the SM vector bosons mix with heavy spin-1 resonances.
The corresponding EW sector is 
contained in one of three cosets: $\SU(5)/\SO(5)$, 
$\SU(4)/\Sp(4)$, and $\SU(4)^2/\SU(4)$ depending
on whether the hyperfermions are in a real, pseudo-real or complex irreducible representation (irrep) of $G_\mathrm{HC}$, 
respectively. 
It turns out that in all three cosets this mixing is the same\footnote{This was shown in ref.~\cite{Caliri:2024jdk} for the real and pseudo-real case, and after publication of that work we found that with a change of basis, the mixing matrices in $\SU(4)^2/\SU(4)$ can be turned into the same form as well.}.
The states mixing with the SM gauge bosons $B_\mu$ and $W_\mu$ are
\begin{align}
    a_\mu &\in (\mathbf 3, \mathbf 1), \quad v_{1\mu} \in (\mathbf 3, \mathbf 1), \quad v_{2\mu} \in (\mathbf 1, \mathbf 3) \qquad \text{of } \SU(2)_L\times \SU(2)_R,
\end{align}
yielding charge eigenstates 
\begin{align}
    a_\mu^\pm, a_\mu^0, \quad v_{1\mu}^\pm , v_{1\mu}^0, \quad v_{2\mu}^\pm, v_{2\mu}^0.
\end{align}
In a slight abuse of language we call $v_{1\mu}$ and $v_{2\mu}$ vector states and $a_{\mu}$ an axial-vector state. 
The charged states mix with the $\hat W^\pm$ and the neutral states with the $\hat Z$, where the hat indicates that these are not yet the physical $W^\pm$ and $Z$.
Accordingly, they couple with pre-mixing gauge couplings $\hat g, \hat g'$.
To relate them to the physical $g$ and $g'$, we rotate to the mass eigenbasis.

The mass matrices for these states have been calculated in \cite{BuarqueFranzosi:2016ooy,Caliri:2024jdk} and are reproduced in \cref{app:model} for completeness.
The eigenvector of the massless photon can be obtained analytically,
\begin{align}
    A_\mu &= \frac{e}{\hat g} \tilde W_\mu^3 + \frac{e}{ \hat g'} B_\mu + \frac{\sqrt{2} e}{\gt} v_{1\mu}^0 
\quad \text{ with } \quad
    \frac{1}{e^2} = \frac{1}{ \hat g^2} + \frac{1}{ \hat g'^2} + \frac{2}{\Tilde{g}^2} ,
\label{eq:photon}
\end{align}
where $\tilde g$ is the gauge coupling of the composite spin-1 states.
For the full spectrum we diagonalise the mass matrices numerically,
\begin{align}
    \mqty( \tilde W^+_\mu \\ a^+_\mu \\ v_{1\mu}^+ \\ v_{2\mu}^+ ) = \mathcal{C} \mqty( W^+_\mu \\ A^+_\mu \\ V_{1\mu}^+ \\ V_{2\mu}^+ ) = \mathcal{C} V_\mu^+ \,, \qquad \mqty( B_\mu \\ \tilde W^3_\mu \\ a^0_\mu \\ v_{1\mu}^0 \\ v_{2\mu}^0 ) = \mathcal{N} \mqty( A_\mu \\ Z_\mu \\ A^0_\mu \\ V_{1\mu}^0 \\ V_{2\mu}^0 ) = \mathcal{N} V_\mu^0 \, \label{eq:mass-eigenstates-basis}.
\end{align}
We denote the mass eigenstates with capital letters $V_\mu^+$ and $V_\mu^0$.
The orthogonal rotation matrices $\mathcal{C}$ and $\mathcal{N}$ relate the gauge to the physical states, with $\mathcal C_{1i}$ describing the mixing of the charged BSM vectors to the $W$ and $\mathcal N_{1i}$ and $\mathcal N_{2i}$ the mixing of the neutral BSM vectors to $A_\mu$ and $Z_\mu$, respectively.
These mixings induce couplings of the corresponding spin-1
resonances to the SM quarks and leptons:
\begin{align}
    \mathcal{L}_\mathrm{CC} &= \frac{\hat g}{\sqrt{2}} \sum_{i,f,f'} \mathcal{C}_{1i} \overline{\psi}_f \slashed{V}^+_i P_\mathrm{L} (V_{\tiny \mathrm{CKM}})_{ff'}\psi_{f'} + \hc  \label{eq:LagCC}\\
    \mathcal{L}_\mathrm{NC} &=  \sum_{i,f} \overline{\psi}_f \slashed{V}^0_i \left( g_{\mathrm{L}i}^f P_\mathrm{L} + g_{\mathrm{R}i}^f P_\mathrm{R} \right) \psi_f  \label{eq:LagNC}
\end{align}
For a fermion $f$ with weak isospin $T_f^3$ and hypercharge $Y_{f,L/R}$, the chiral couplings are given by
\begin{align}
g_{\mathrm{L}i}^f = \hat g T^3_f \mathcal{N}_{2i} + \hat g' Y_{f,L} \mathcal{N}_{1i}  \quad \text{ and } \quad
g_{\mathrm{R}i}^f = \hat g' Y_{f,R} \mathcal{N}_{1i}.
\end{align}
From inserting the rotation into \cref{eq:photon} we find the relation between the $\hat g,\hat g'$ with the physical gauge couplings,
\begin{align}\label{eq:deficouplings}
    \hat g' \mathcal{N}_{11} = \hat g \mathcal{N}_{21} = e = g' \cos\theta_W = g \sin \theta_W,
\end{align}
with $\theta_W$ the Weinberg angle. 
This implies that the coupling of the $W$ to two fermions is modified with respect to the SM, since $\hat g \mathcal C_{11} \ne g$.
We have checked numerically that the deviation is small. 
Furthermore, in models with partial compositeness the couplings of the third-generation quarks to the spin-1 resonances get additional contributions, which however do not enter the VBS processes.
We refer to ref.~\cite{Caliri:2024jdk} for the corresponding details.

The mixing in the spin-1 sector implies further modifications of several SM couplings as well as new interactions that play an important role for VBS processes.
The trilinear and quartic gauge couplings of the SM bosons are modified and vertices of type
\begin{align}
    \mathcal V {\mathcal V} \tilde{\mathcal V}, \, \mathcal V \tilde{\mathcal V} \tilde{\mathcal V} ,\, \tilde{\mathcal V}\tilde{\mathcal V}\tilde{\mathcal V} ; \quad \tilde{\mathcal V} {\mathcal V} {\mathcal V} {\mathcal V},\, \tilde{\mathcal V} \tilde{\mathcal V} \mathcal V \mathcal V
\end{align}
are added, where $\mathcal V = W,Z,\gamma$ and $\tilde{\mathcal V}$ denotes a spin-1 resonance mixing with the SM vector bosons. 
Quartic couplings $\tilde{\mathcal V} \tilde{\mathcal V} \tilde{\mathcal V} \mathcal V$, $\tilde{\mathcal V}\tilde{\mathcal V}\tilde{\mathcal V}\tilde{\mathcal V}$ with three and four BSM vectors are also present but do not contribute to VBS processes at tree-level.
In addition, there are couplings
\begin{align}
    \mathcal V \tilde{\mathcal V} h
\end{align}
of a single spin-1 resonance, an EW vector boson, and the Higgs boson. 
Analogous couplings in which $h$ is replaced by another pNGBs $\pi$ do not exist, as we assume that the Higgs is the only field contributing to EWSB.
Couplings of $h$ to only BSM vectors are also present but not relevant for this work. 
The couplings of two EW vector bosons and the Higgs boson are modified in the composite Higgs model.
We do not give the corresponding expressions here, as they are quite lengthy, and refer to ref.~\cite{Caliri:2024jdk} for details instead.

The model contains several parameters, which can be reduced to a set of five independent input parameters \cite{Caliri:2024jdk}.
We choose the mass parameters $M_V$ and $M_A$ of the vector and axial vector states, which correspond to the physical masses apart from small mixing effects.
Next, we choose the couplings $\tilde g$ and $g_{V\pi\pi}$, respectively the gauge couplings of the spin-1 resonances and the coupling of two pNGBs to a spin-1 resonance, the analogue of $g_{\rho\pi\pi}$ in QCD.
The value of $\tilde g$ strongly determines the strength of the mixing and therefore the differences in VBS processes relative to the SM.
As a final input parameter we take the pNGB decay constant $f_\pi$.
The gauge couplings are fixed through \cref{eq:photon} and we require the mass matrix $\mathcal M_{N}^2$ for the neutral states to have an eigenvalue corresponding to the $Z$ mass.

\subsection{Benchmark scenarios}\label{sec:bm}

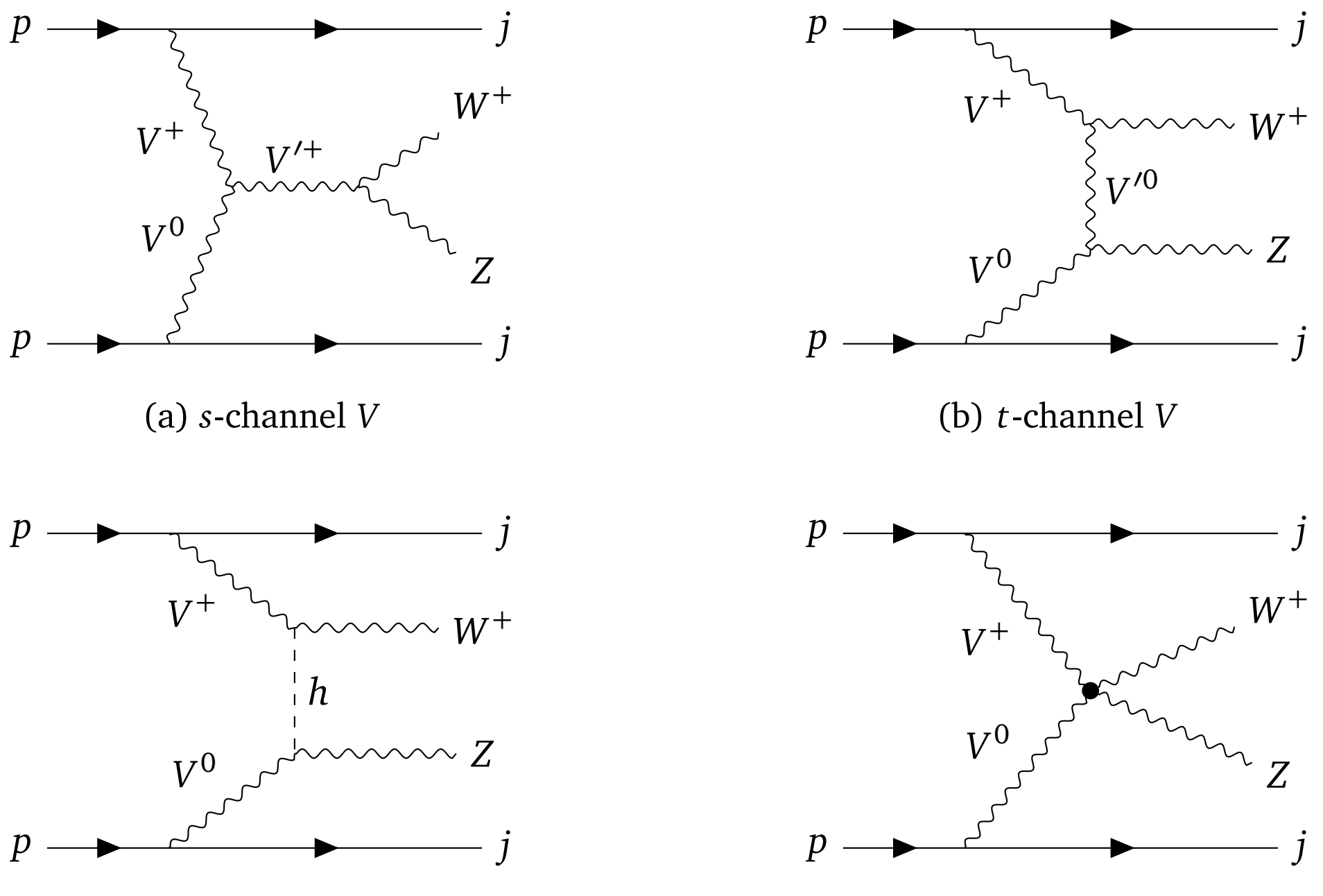
\begin{figure}[t]
  \centering
  \begin{subfigure}{0.45\textwidth}
    \centering
    \begin{tikzpicture}
      \begin{feynman}
        \vertex (i1) at (0, 3) {\(p\)};
        \vertex (v1) at (1.4, 3);
        \vertex (f1) at (4.6, 3) {\(j\)};
        \vertex (i2) at (0, 0) {\(p\)};
        \vertex (v2) at (1.4, 0);
        \vertex (f2) at (4.6, 0) {\(j\)};
        \vertex (s1) at (2.0, 1.5);
        \vertex (s2) at (3.2, 1.5);
        \vertex (w) at (4.4, 2.3) {\(W^+\)};
        \vertex (z) at (4.4, 0.7) {\(Z\)};
        \diagram* {
          (i1) -- [fermion] (v1) -- [fermion] (f1),
          (i2) -- [fermion] (v2) -- [fermion] (f2),
          (v1) -- [boson, edge label'=\(V^+\)] (s1),
          (v2) -- [boson, edge label=\(V^0\)] (s1),
          (s1) -- [boson, edge label=\(V'^+\)] (s2),
          (s2) -- [boson] (w),
          (s2) -- [boson] (z),
        };
      \end{feynman}
    \end{tikzpicture}
    \caption{$s$-channel $V$}
    \label{fig:vbs_s}
  \end{subfigure}
  \qquad
  \begin{subfigure}{0.45\textwidth}
    \centering
    \begin{tikzpicture}
      \begin{feynman}
        \vertex (i1) at (0, 3) {\(p\)};
        \vertex (v1) at (1.4, 3);
        \vertex (f1) at (4.6, 3) {\(j\)};
        \vertex (i2) at (0, 0) {\(p\)};
        \vertex (v2) at (1.4, 0);
        \vertex (f2) at (4.6, 0) {\(j\)};
        \vertex (t1) at (2.6, 2.1);
        \vertex (t2) at (2.6, 0.9);
        \vertex (z) at (4.4, 2.1) {\(W^+\)};
        \vertex (w) at (4.4, 0.9) {\(Z\)};
        \diagram* {
          (i1) -- [fermion] (v1) -- [fermion] (f1),
          (i2) -- [fermion] (v2) -- [fermion] (f2),
          (v1) -- [boson, edge label'=\(V^+\)] (t1),
          (v2) -- [boson, edge label=\(V^0\)] (t2),
          (t1) -- [boson, edge label=\(V'^0\)] (t2),
          (t1) -- [boson] (z),
          (t2) -- [boson] (w),
        };
      \end{feynman}
    \end{tikzpicture}
    \caption{$t$-channel $V$}
    \label{fig:vbs_t}
  \end{subfigure}
  \\[1.5em]
  \begin{subfigure}{0.45\textwidth}
    \centering
    \begin{tikzpicture}
      \begin{feynman}
        \vertex (i1) at (0, 3) {\(p\)};
        \vertex (v1) at (1.4, 3);
        \vertex (f1) at (4.6, 3) {\(j\)};
        \vertex (i2) at (0, 0) {\(p\)};
        \vertex (v2) at (1.4, 0);
        \vertex (f2) at (4.6, 0) {\(j\)};
        \vertex (t1) at (2.6, 2.1);
        \vertex (t2) at (2.6, 0.9);
        \vertex (w) at (4.4, 2.1) {\(W^+\)};
        \vertex (z) at (4.4, 0.9) {\(Z\)};
        \diagram* {
          (i1) -- [fermion] (v1) -- [fermion] (f1),
          (i2) -- [fermion] (v2) -- [fermion] (f2),
          (v1) -- [boson, edge label'=\(V^+\)] (t1),
          (v2) -- [boson, edge label=\(V^0\)] (t2),
          (t1) -- [scalar, edge label=\(h\)] (t2),
          (t1) -- [boson] (w),
          (t2) -- [boson] (z),
        };
      \end{feynman}
    \end{tikzpicture}
    \caption{$t$-channel Higgs}
    \label{fig:vbs_h}
  \end{subfigure}
  \qquad
  \begin{subfigure}{0.45\textwidth}
    \centering
    \begin{tikzpicture}
      \begin{feynman}
        \vertex (i1) at (0, 3) {\(p\)};
        \vertex (v1) at (1.4, 3);
        \vertex (f1) at (4.6, 3) {\(j\)};
        \vertex (i2) at (0, 0) {\(p\)};
        \vertex (v2) at (1.4, 0);
        \vertex (f2) at (4.6, 0) {\(j\)};
        \vertex [dot] (q) at (2.6, 1.5) {};
        \vertex (w) at (4.4, 2.3) {\(W^+\)};
        \vertex (z) at (4.4, 0.7) {\(Z\)};
        \diagram* {
          (i1) -- [fermion] (v1) -- [fermion] (f1),
          (i2) -- [fermion] (v2) -- [fermion] (f2),
          (v1) -- [boson, edge label'=\(V^+\)] (q),
          (v2) -- [boson, edge label=\(V^0\)] (q),
          (q) -- [boson] (w),
          (q) -- [boson] (z),
        };
      \end{feynman}
    \end{tikzpicture}
    \caption{quartic gauge coupling}
    \label{fig:vbs_q}
  \end{subfigure}
  \caption{Feynman diagrams of the vector boson scattering processes for
    $p p \to W^+Z j j$. $V^0$, $V^{(\prime)+}$
    can be either a SM vector boson or a spin-1 resonance. In case of (b) there is also the corresponding $u$-channel, in which $V'{}^0$ is replaced by a $V'{}^+$.}
  \label{fig:vbs_diagrams}
\end{figure}

The discussion above shows that the parameter space is too large to allow for a systematic scan. 
Instead we study a few benchmark points.
Since the axial states only play a subdominant role for VBS, we fix their mass, taking $M_A/M_V = 1.4$ as both Lattice studies \cite{Ayyar:2018zuk,Bennett:2019jzz,Bennett:2019cxd,Bennett:2023qwx} and holographic models using gauge/gravity duality \cite{Erdmenger:2020lvq,Erdmenger:2020flu,Elander:2020nyd,Elander:2021bmt,Erdmenger:2023hkl,Erdmenger:2024dxf} suggest the axial states to be somewhat heavier.
The $g_{V\pi\pi}$ is a composite sector coupling, so we expect it to be large. 
In the following we fix it to $g_{V\pi\pi}=4$.
The symmetry breaking scale is set to $f_\pi = 1$~TeV.
In ref.~\cite{Caliri:2024jdk}, bounds in the $m_V$-$\tilde g$-plane have been derived using collider signatures of $s$-channel production of the spin-1 resonances.
In the following analysis we select benchmark points where we expect sizeable deviations from the SM while evading the bounds from \cite{Caliri:2024jdk}.
To this end, we choose two benchmarks with vector masses of 2~TeV and 2.5~TeV, both with a coupling of $\tilde g = 4$.

Having fixed the benchmarks, the next step is to select the process that is most sensitive to the differences between the SM and the CH model.
In general, large cancellations take place between different contributions in VBS processes due to gauge invariance.
This makes them an ideal test for modifications of triple and quartic gauge couplings as well as the couplings of two vector bosons to the Higgs boson.
The cleanest process would be $pp\to W^+ W^+ jj$ as it has the lowest background, see, e.g.~\cite{BuarqueFranzosi:2021wrv} and references therein. 
However, it turns out that in the model class considered here, the corresponding effects are too small once the constraints from existing LHC data are taken into account. 
This is to a large extent due to the fact that in these models there are only $t$- and $u$-channel contributions due to the additional particles\footnote{At first glance this is somewhat surprising for the $\SU(5)/\SO(5)$ coset which predicts a doubly charged pNGB and a doubly charged spin-1 resonance, both of which could in principle show up in the $s$-channel. 
However, the coupling of the spin-1 resonance to two $W$-bosons vanishes as it is an axial-vector, while the doubly charged scalar has only an anomaly-induced and thus loop-suppressed coupling to two $W$-bosons \cite{Agugliaro:2018vsu,Flacke:2023eil}.}.
We delegate the results for this case to \cref{app:WW}.

The next cleanest process is 
\begin{align}
p p \to W^{\pm}Z jj,
\end{align}
which we take as the target process in this work.
We display the relevant Feynman diagrams of the VBS topology in \cref{fig:vbs_diagrams}, where mediators labelled $V^{0,+}$ can be SM vector bosons or heavy spin-1 resonances.
The contributions from \cref{fig:vbs_s} are of particular importance as they contain an $s$-channel exchange with a heavy mediator. 
This gives the dominant contribution to changes for the production of longitudinally polarised vector bosons.
Taking into account all Feynman diagrams leading to $W^\pm Zjj$ production, including QCD-induced diagrams, there are 24064 Feynman diagrams contributing in the SM while the CH model has 146464 diagrams. 
Here we only consider CKM mixing of the first two generations.

\section{Event generation and selection cuts}
\label{sec:vbs}

In a CH model, SM vertices of the gauge bosons are modified and new states can appear as propagators.
We expect this to modify the kinematics of the produced $W$ and $Z$ bosons as well as their decay products.
To verify this, we perform a collider study modelled after a recent experimental VBS study by the ATLAS collaboration \cite{ATLAS:2024ini}.

We implement the model described in the previous section in the \textsc{FeynRules}  \cite{Alloul:2013bka}  framework to produce a \textsc{UFO} \cite{Degrande:2011ua} library\footnote{The corresponding UFO file can be obtained on request from the authors.}.
For the SM, we use the implementation \cite{fukssmufo} with CKM elements coupling the third to the first two generations set to zero.
We generate hard scattering events for 
\begin{equation}
    pp \to W^\pm Z jj \to \left( \ell^\pm \nu \right) \left( \ell^{\prime +} \ell^{\prime -} \right) jj,
\end{equation}
where both $W^+$ and $W^-$ are taken into account.
Event generation is performed within \textsc{MadGraph5\_aMC@NLO} \cite{Alwall:2014hca}  version 3.6.5 at a centre-of-mass energy of $\sqrt s = 13$~TeV.
We use the \textsc{NNPDF} 2.3 PDF set \cite{Ball:2012cx} provided by \textsc{LHAPDF}  \cite{Buckley:2014ana}.
We implemented a custom scale choice following \cite{ATLAS:2023sua}, setting $\mu_R = \mu_F = (\pT^{j_1} \pT^{j_2})^{1/2}$. 
To improve the acceptance of the analysis described below, we enforce $M_{jj}>400$~GeV and $\pT^j > 20$~GeV at the generator level, with stricter cuts towards the VBS topology applied later on.
We focus on the leptonic decays of the SM bosons and use \textsc{MadSpin} \cite{Artoisenet:2012st} to decay them to leptons $\ell = e,\mu,\tau$ and neutrinos. 
The partonic events are then passed to \textsc{Pythia8} \cite{Sjostrand:2014zea} for showering and hadronisation, at which point also the $\tau$ leptons from the gauge bosons are further decayed to jets or leptons, producing a \texttt{HepMC} file \cite{Dobbs:2001ck}.

To assess the impact of polarisation, we generate both an unpolarised event sample (unpol) and a sample where both vector bosons are longitudinally polarised (LL) for each parameter point.
To ensure sufficient statistics, we generate $10^6$ events for each sample.
We only perform a leading-order calculation as higher-order corrections are not known in case of composite Higgs models. The cross section after generator-level cuts is $\sigma_{\text{SM}} = 81.20$~fb (3.96~fb) for the unpolarised (LL) sample.
The additional spin-1 states change the total cross section only at the permille level in case of our benchmark scenarios.

\begin{table}
  \centering
  \begin{tabular}{lll}
    \toprule
    Object & Baseline & Signal \\
    \midrule
    Electrons & $\pT > 4.5\GeV$, $|\eta| < 2.47$, & $\pT > 27\GeV$, $|\eta| < 1.37$ \\
              & excl.\ $1.37 < |\eta| < 1.52$ & \\
    Muons     & $\pT > 3\GeV$, $|\eta| < 2.7$, & $\pT > 27\GeV$, $|\eta| < 2.5$ \\
    Jets      & \multicolumn{2}{c}{$\pT > 25\GeV$, $|\eta| < 4.5$} \\
    $b$-jets  & \multicolumn{2}{c}{$\pT > 20\GeV$, $|\eta| < 2.5$} \\
    \midrule
    \multicolumn{3}{l}{remove jets with $\Delta R(j,\ell) < 0.2$ to any baseline lepton} \\
    \multicolumn{3}{l}{remove baseline leptons with $\Delta R(\ell,j) < 0.4$ to any remaining jet} \\
    \bottomrule
  \end{tabular}
    \caption{Object definitions of leptons and jets.
    Overlap removal is applied in the order listed; $b$-jets are not overlap-removed.}
  \label{tab:objects}
\end{table}

\begin{table}
  \centering
  \begin{tabular}{rl}
    \toprule
    & $WZjj$\\
    \midrule
    Leptons & $N_\ell= 3$, not all leptons of the same charge \\
    $Z$ boson &
    SFOS pair with $|m_{\ell\ell} - \mZ| < 15\GeV$ \\
    Jets
     &$N_j \geq 2$, $\pT(j_1) > 65\GeV$, $\pT(j_2) > 35\GeV$ \\
    $MET$         &  $> 30\GeV$ \\
    $b$-jet veto   & $N_b = 0$ \\
    VBS topology
      & $m_{j_1 j_2} > 500\GeV$, 
       $|\Delta y_{j_1 j_2}| > 2$ \\
    \bottomrule
  \end{tabular}
    \caption{Event selection defining the VBS signal region. A same flavour opposite sign (SFOS) pair of leptons is required to have an invariant mass consistent with $Z\to \ell^+ \ell^-$. The jets $j_{1,2}$ denote the two hardest signal jets.}
  \label{tab:cuts}
\end{table}

The generated events are then passed through an analysis designed to isolate the VBS topology along the lines of ref.~\cite{ATLAS:2024ini}.
We implement it in \texttt{MadAnalysis5} \cite{Conte:2012fm,Conte:2014zja,Conte:2018vmg}.
Detector simulation is performed with the simplified fast simulation (SFS) \cite{Araz:2020lnp} framework. 
We use the default ATLAS settings in SFS.
We define baseline muons (electrons) by requiring pseudo rapidity $|\eta| < 2.7$ ($|\eta| < 2.47$ excluding $1.37 < |\eta| < 1.52$) and transverse momentum $\pT > 3$~GeV ($\pT > 4.5$~GeV).
The baseline leptons are used for overlap removal with jets as detailed below.
Signal leptons $\ell = e,\mu$ are instead required to have $\pT>27$~GeV and $|\eta| < 1.37$ for electrons and $|\eta|<2.5$ for muons. 
Jets are clustered with the anti-$k_T$ algorithm \cite{Cacciari:2008gp} implemented in \textsc{FastJet} \cite{Cacciari:2011ma} using a jet radius of $R=0.4$.
The signal light jets, which are candidates for the VBS tagging jets, satisfy $|\eta|<4.5$ and $\pT>25$~GeV.
The signal $b$-jets, used later on to enforce a $b$-jet veto, have $|\eta|<2.5$ and $\pT>20$~GeV.
A $\pT$- and $\eta$-dependent $b$-tagging efficiency is used \cite{Araz:2020lnp}.
Among these jets, we removed the ones within a cone of $\Delta R = \sqrt{(\Delta \phi)^2 + (\Delta \eta)^2}=0.2$ of a lepton, and the baseline leptons within $\Delta R$ of $0.4$ of a remaining jet. 
Finally, baseline electrons (muons) with $|\eta|< 1.37$ and $\pT > 27 \GeV$ ($|\eta|< 2.5$ and $\pT > 27 \GeV$) are taken as signal electrons (muons). 
The object definitions are summarised in \cref{tab:objects}.

Next, we apply selection cuts to isolate the VBS topology events.
We require three signal leptons with charges $\pm\pm\mp$ matching the fully leptonic $WZ$ decays.
Two leptons must form a same-flavour pair with opposite sign with an invariant mass consistent with a decay from a $Z$-boson\footnote{This cut eliminates almost all events with $Z \to \tau^+ \tau^-$. To save computational resources, we therefore omitted this decay in the event generation.}, $|m_{\ell\ell} - m_Z| < 15$~GeV.
Events must contain at least two signal jets and no b-jets.
We demand the two leading jets to satisfy $\pT(j_1) > 65$~GeV and $\pT(j_2) > 35$~GeV.
The leptonic decay of the W boson produces a neutrino. This is taken into account by requiring some missing transverse energy to be present, $MET > 30$~GeV. 
Finally, to narrow in on the VBS topology with backward-forward jet production, a large dijet invariant mass and large rapidity separation are required, $m_{j_1j_2} > 500$~GeV and $|\Delta y| = |y(j_1) - y(j_2)| > 2$. 
The event selection cuts are summarised in \cref{tab:cuts}.
The unpolarised and polarised sample have efficiencies of 2.33\% and 2.17\%, respectively.
Note that for this sample, generator-level cuts on $m_{jj}$ and $\pT^j$ were enforced.

\section{Results}
\label{sec:results}

\begin{figure}
    \centering
        \includegraphics[width=0.45\linewidth]{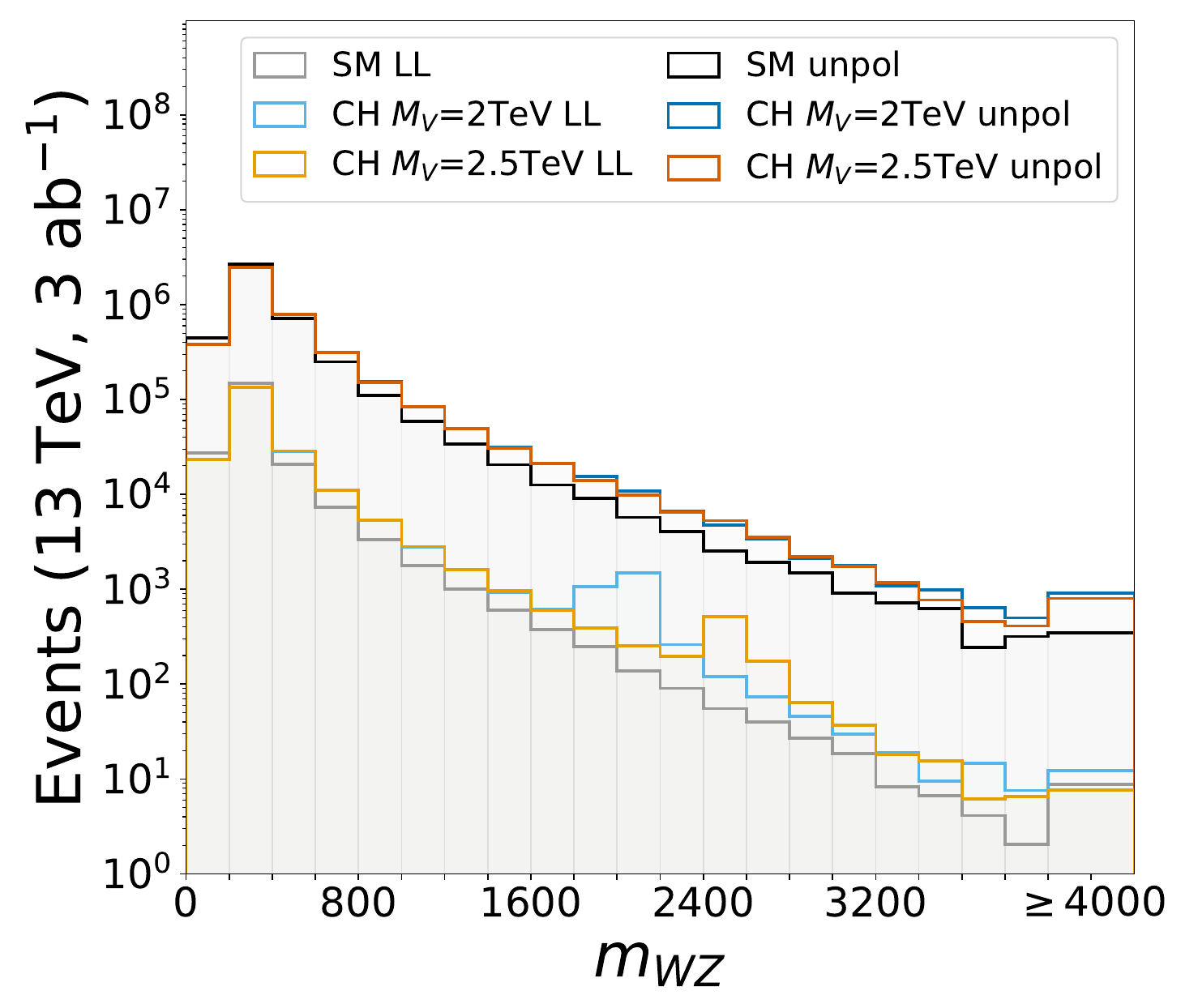}
        \includegraphics[width=0.45\linewidth]{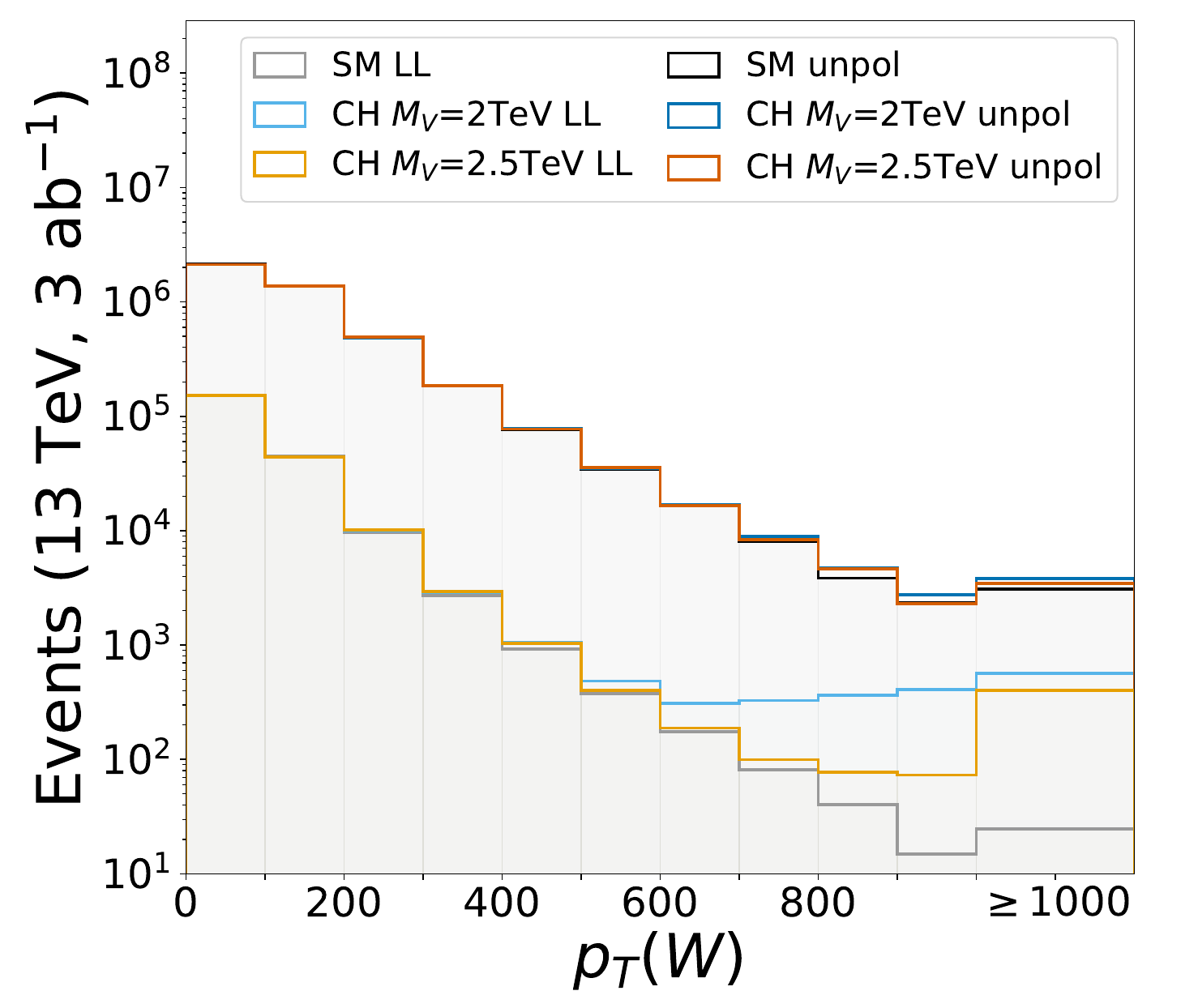}
    \caption{Truth-level distributions of the invariant mass and transverse
    momentum of the $W^\pm Z$ system. The black/grey
    lines correspond to the SM; the blue (orange) ones to CH with
    $M_V = 2$~TeV ($M_V = 2.5$~TeV). Dark lines show the unpolarised samples, and light lines the
    LL samples. The last bin is an overflow bin.}
    \label{fig:wz_truth}
\end{figure}

\begin{figure}[t]
    \centering
        \includegraphics[width=0.45\linewidth]{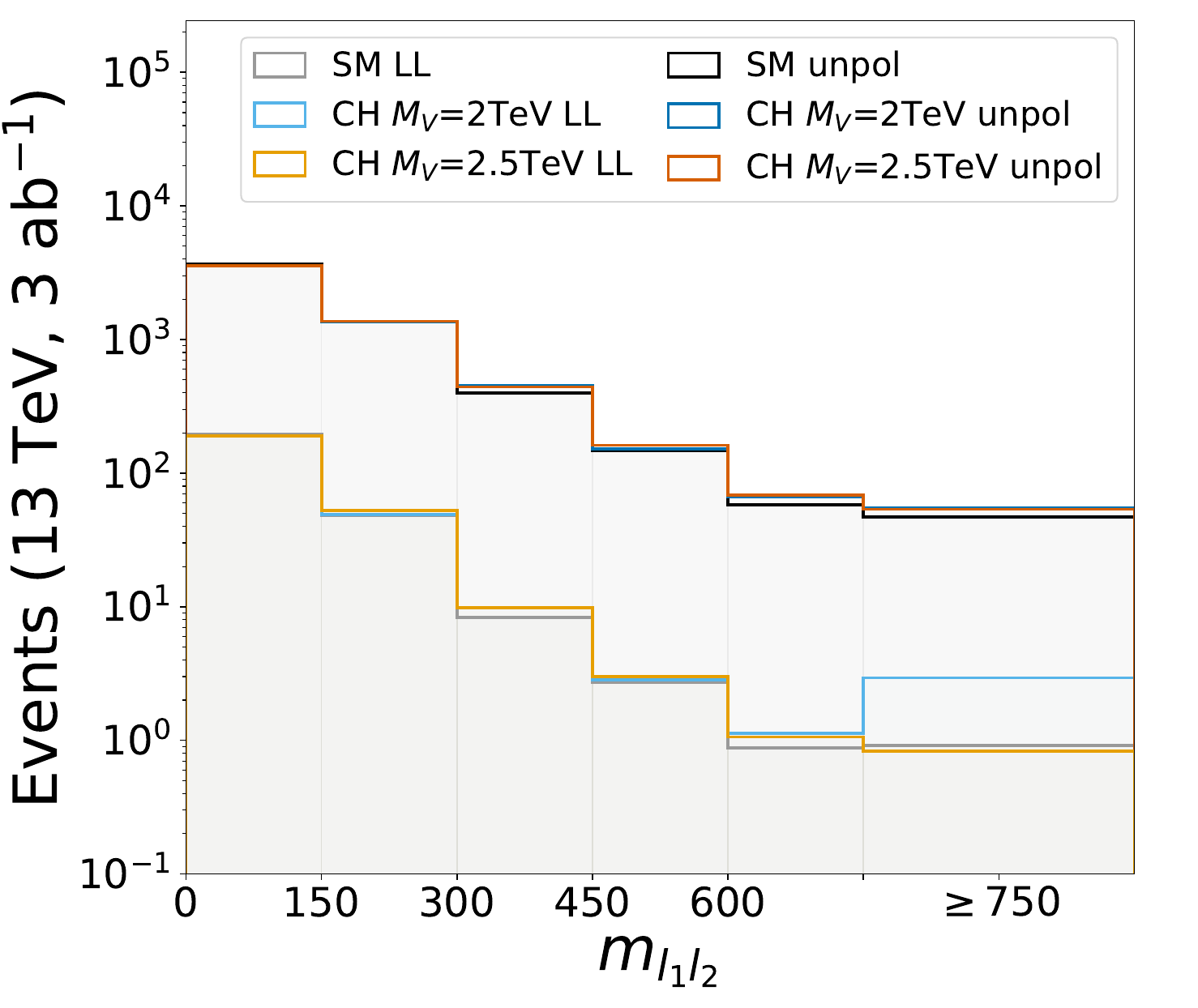}
        \includegraphics[width=0.45\linewidth]{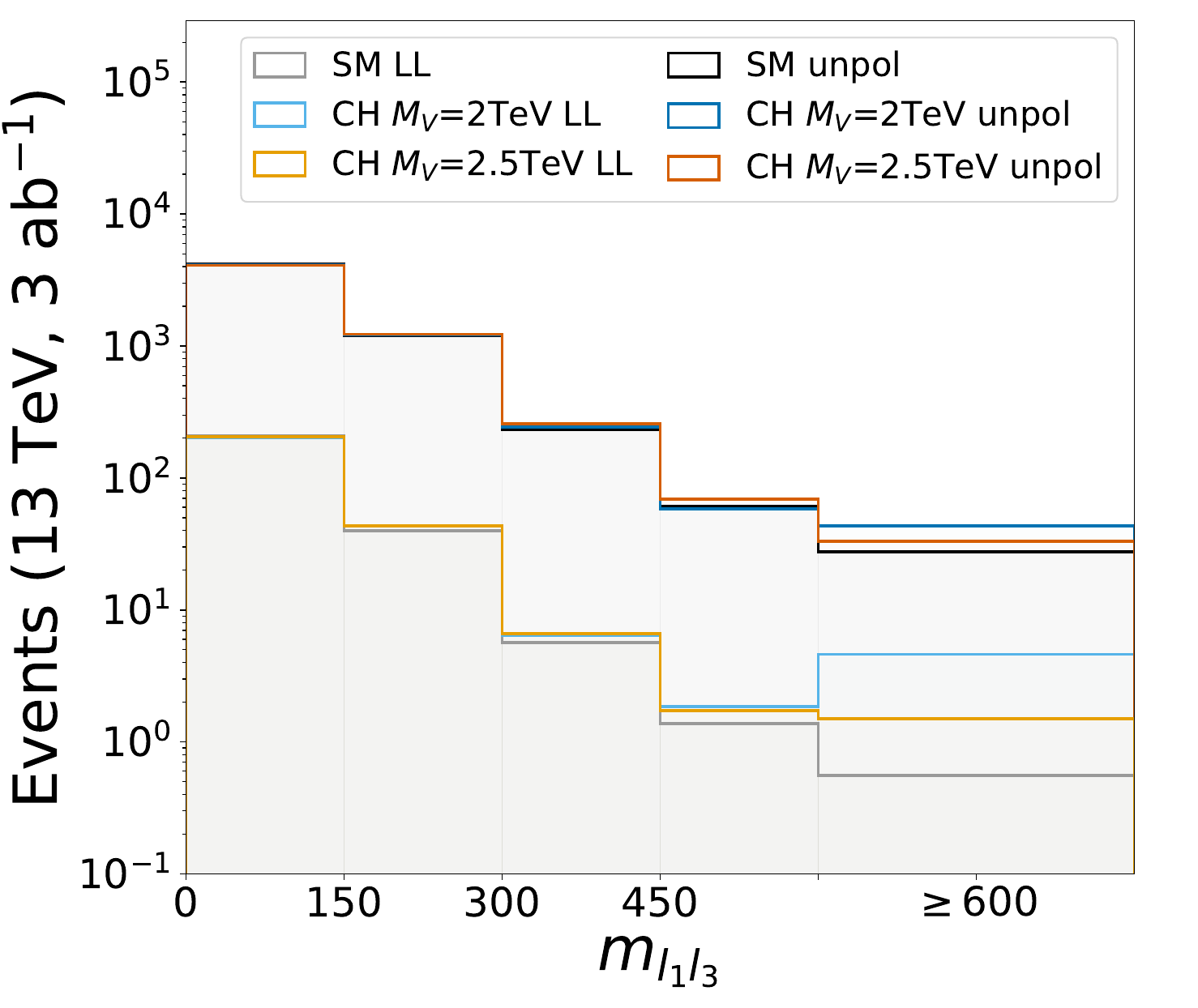}
        
        \includegraphics[width=0.45\linewidth]{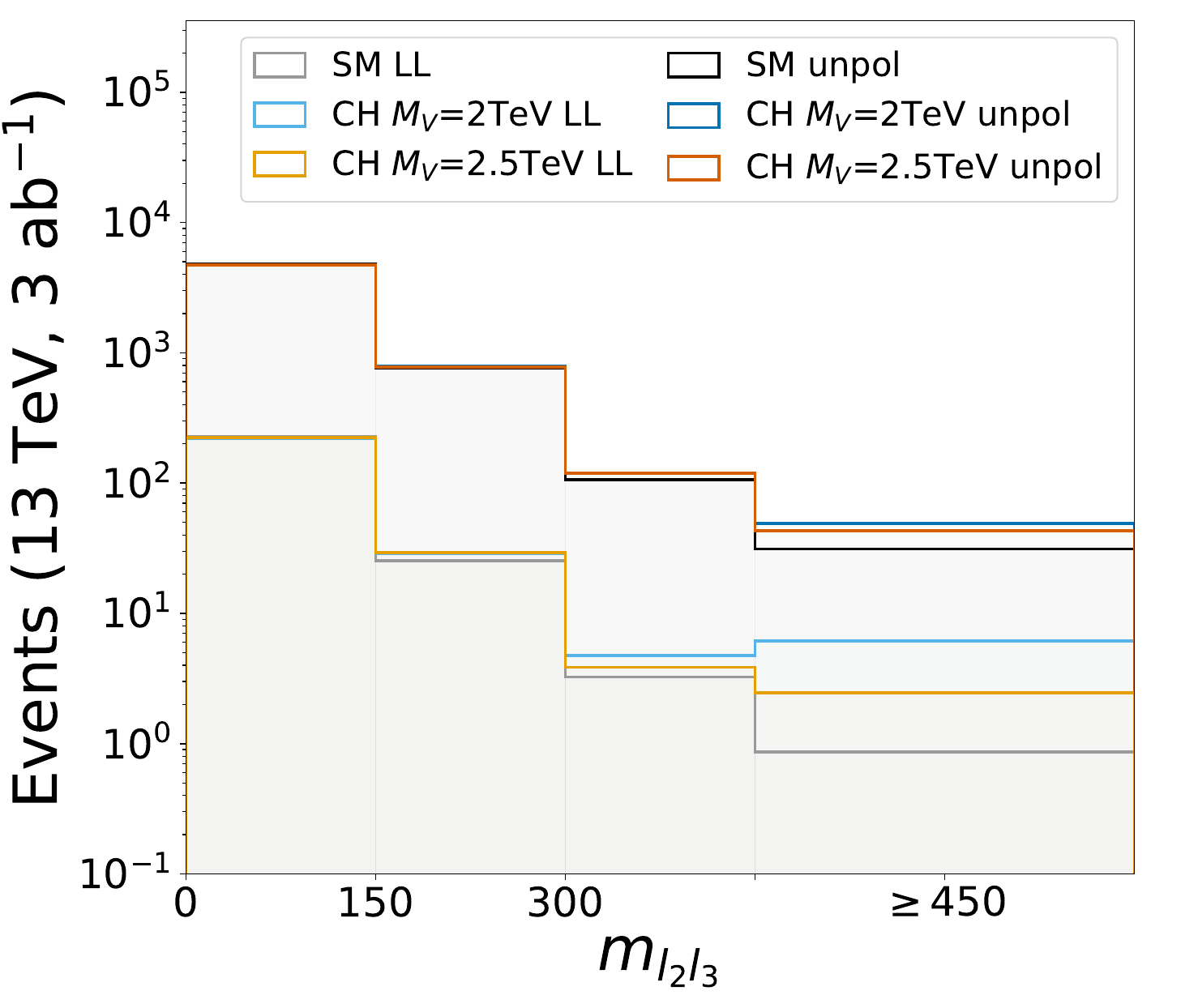}
        \includegraphics[width=0.45\linewidth]{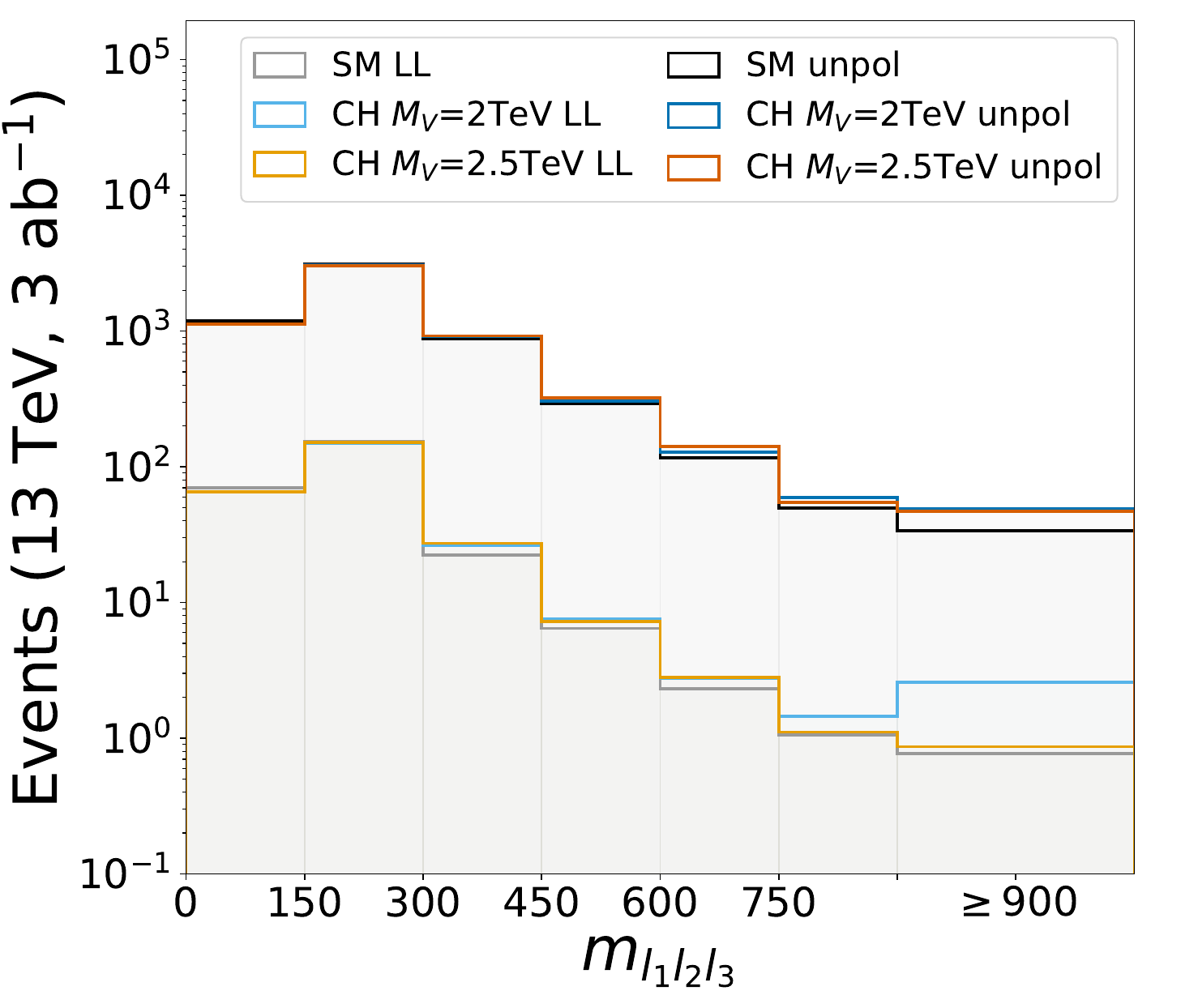}
    \caption{Distributions of the dilepton and
    trilepton invariant masses in $pp \to W^\pm Z jj \to 3\ell\,\nu\,jj$ after the full event
    selection. The black/grey lines correspond to the SM; the blue (orange) one to CH with $M_V = 2$~TeV ($M_V = 2.5$~TeV),
     all with $g_{V\pi\pi} = \tilde g = 4$ and $f = 1$~TeV. Dark
    lines show the unpolarised samples, and light lines the samples with both
    bosons longitudinally polarised (LL). Each curve is normalised to the
    expected event yield at $3~\text{ab}^{-1}$ and  $\sqrt{s} = 13$~TeV.}
    \label{fig:masses}
\end{figure}

In this section we compare the distributions of several kinematic variables within the SM and the composite Higgs model, focusing on the fully leptonic final state of $pp\to W^\pm Zjj$.
As described in \cref{sec:bm}, we study the following two parameter points of the CH model:
\begin{equation}
  M_V = 2~\text{TeV and } 2.5~\text{TeV},
  \qquad g_{V\pi\pi} = 4, \quad \tilde{g} = 4, \quad f = 1~\text{TeV}, \quad M_A/M_V=1.4
  \label{eq:benjmark}
\end{equation}
for unpolarised $W^{\pm}Zjj$ and polarised $W^{\pm}_L Z_L jj$ production.
As a first step we show parton-level distributions of $\pT$ of the $W$ boson and the invariant mass of the $W$ and $Z$ in \cref{fig:wz_truth}, where we require that $m_{jj} > 500$~GeV and for the jets the same cuts of \cref{tab:cuts}.
The distributions are normalised to the expected number of events with a luminosity of $3$~ab$^{-1}$.
The last bin collects the overflow. 
For the unpolarised case, the CH events are lightly enhanced over the SM at large $M_{WZ}$ while no difference is visible for $\pT^W$.
In the polarised sample on the other hand, we can identify peaks in $m_{WZ}$ at the respective vector masses.
These are due to the $V^+$ resonance in the $s$-channel of \cref{fig:vbs_s}.
The resonances have a width of a few hundred GeV, but still $\Gamma/M < 0.1$. 
At large transverse momentum, the longitudinally polarised $W$ is much harder than in the SM. 

In composite Higgs models one expects the main effects to occur in final states mediated by longitudinally polarised vector bosons. 
This is an immediate consequence of the Goldstone equivalence theorem \cite{Cornwall:1974km,Gounaris:1986cr,Chanowitz:1985hj}
which states that at high energies the scattering of the longitudinal components is equivalent to the scattering of the corresponding Nambu-Goldstone bosons.
The effect is in particular pronounced if there are $s$-channel resonances, like the $V^+$ in our case, needed for the unitarisation of the cross section.
The cross section for longitudinally polarised vector bosons is only a small fraction of the total cross section as can be seen from \cref{fig:wz_truth}.
Thus, also at the reconstructed level we expect minor differences for the unpolarised sample while the deviations of the $\pT$ of the $W$ also propagate to the leptons.

\begin{figure}[t]
    \centering
        \includegraphics[width=0.45\linewidth]{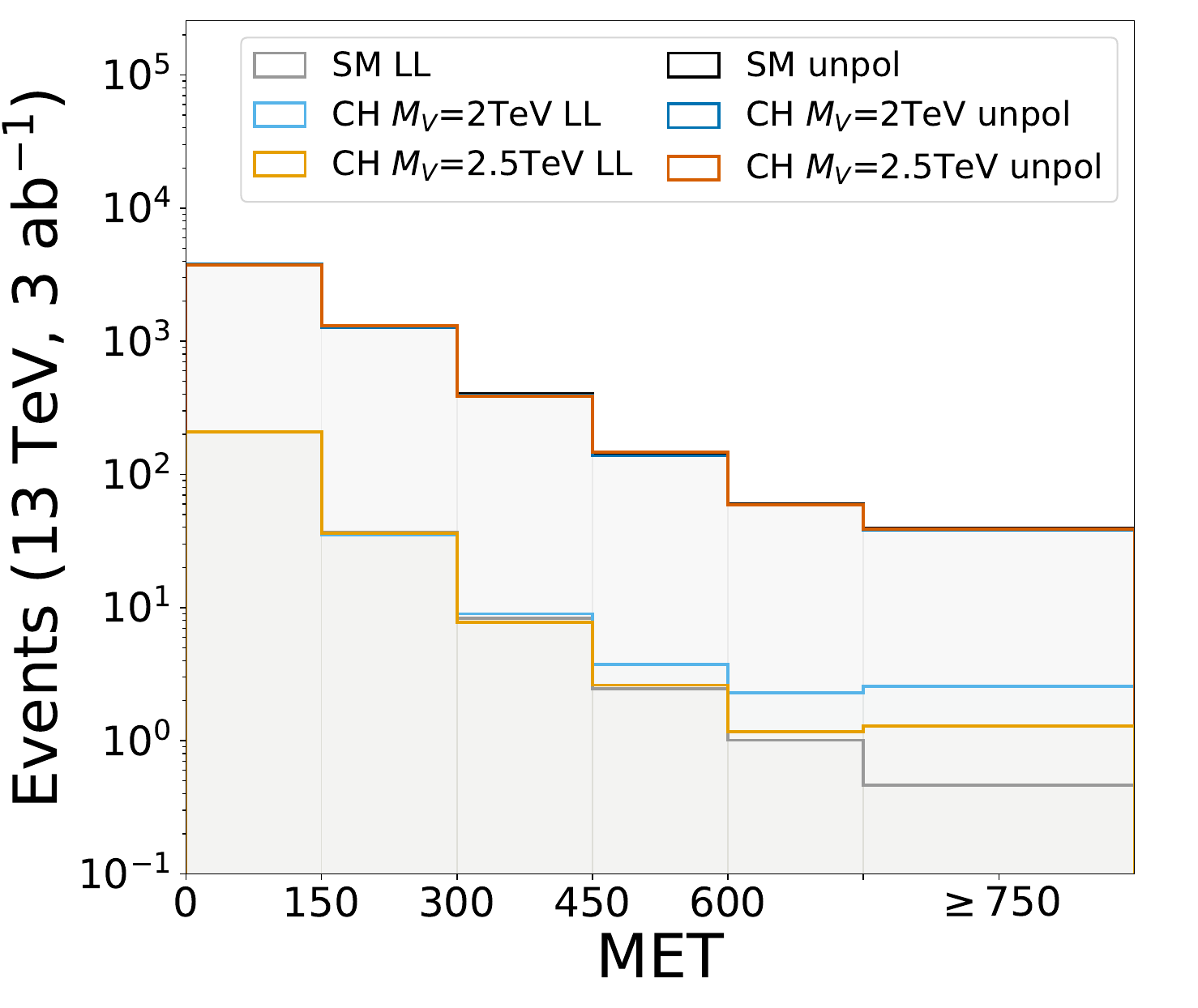}
        \includegraphics[width=0.45\linewidth]{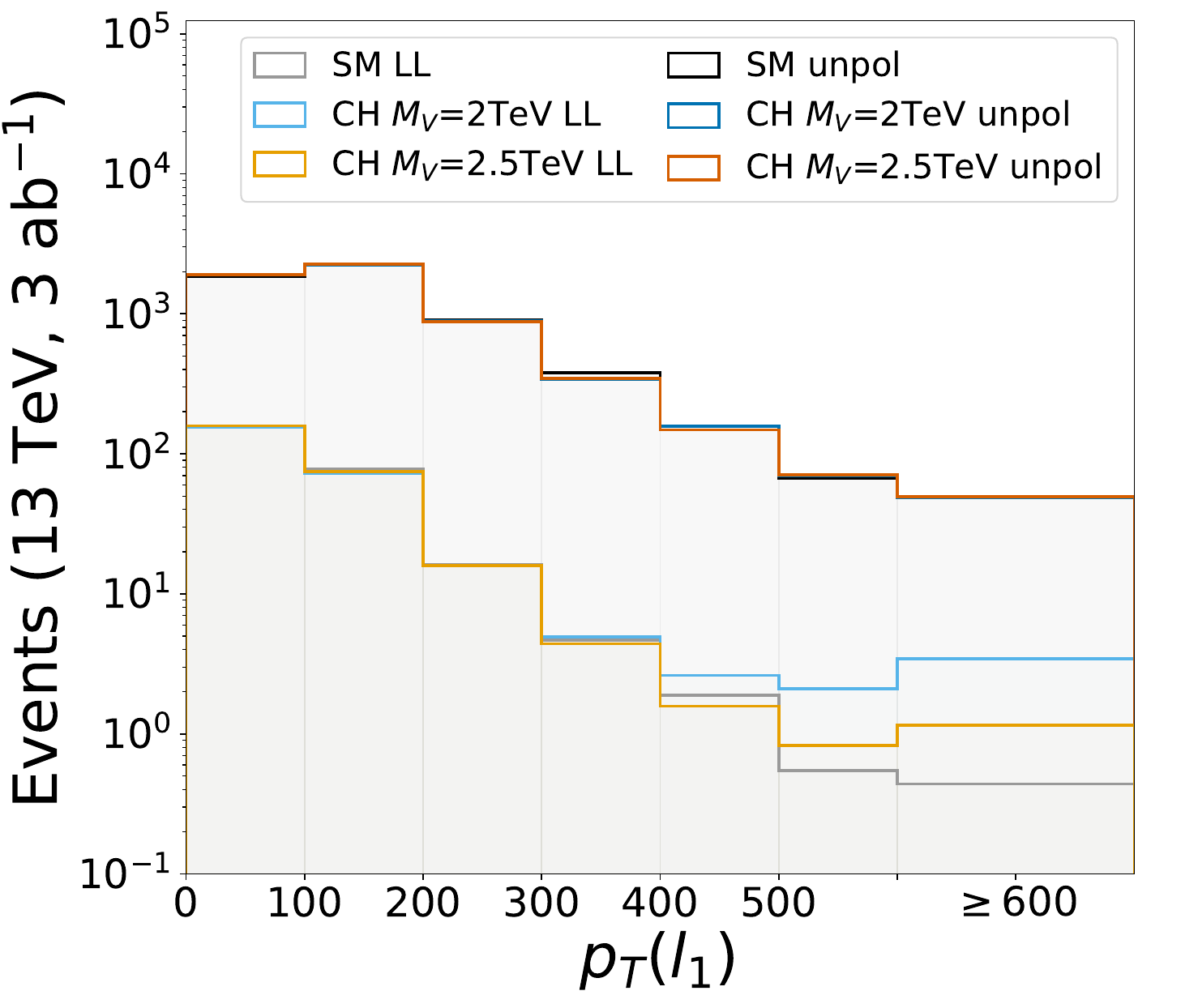}
  
        \includegraphics[width=0.45\linewidth]{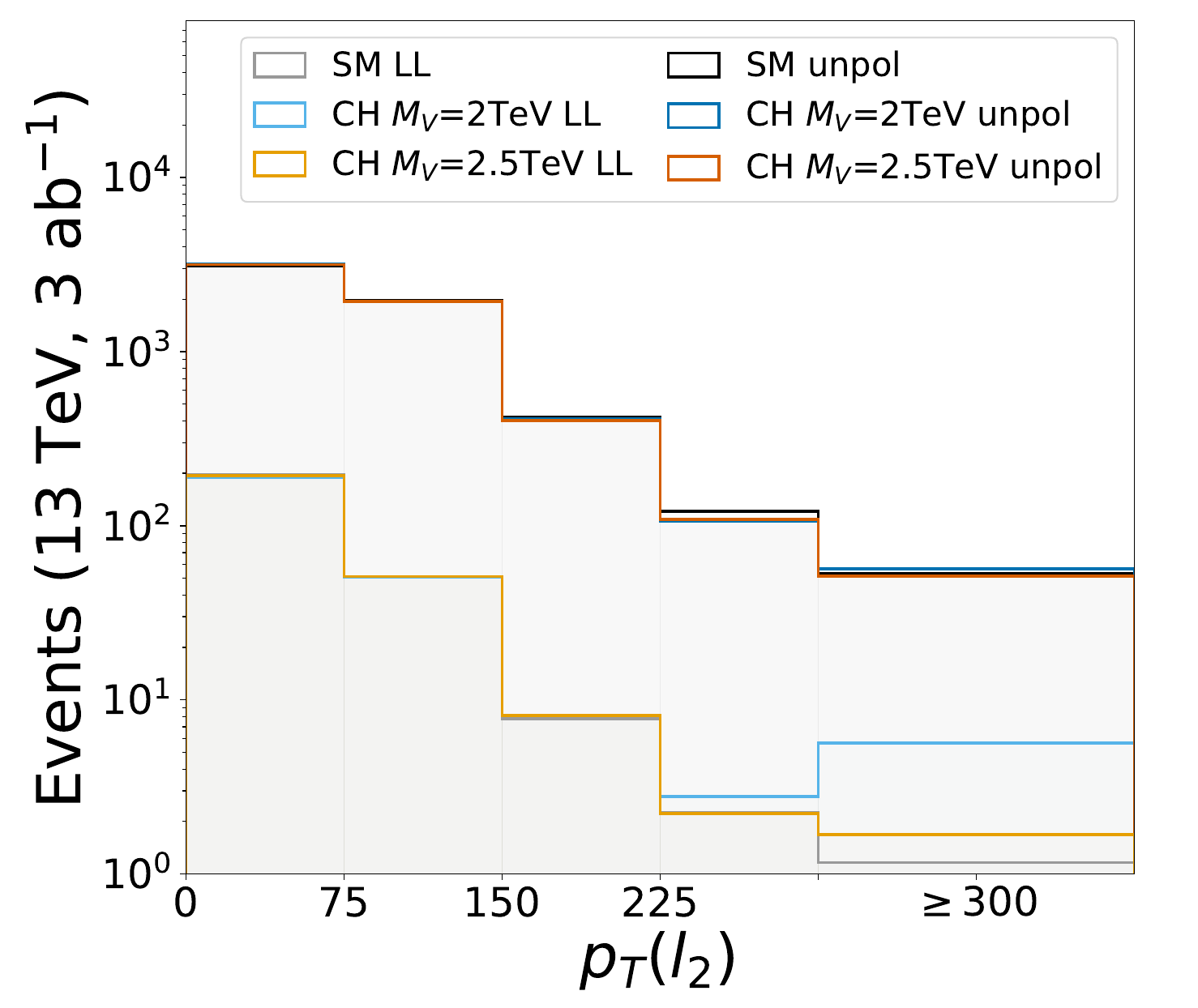}
        \includegraphics[width=0.45\linewidth]{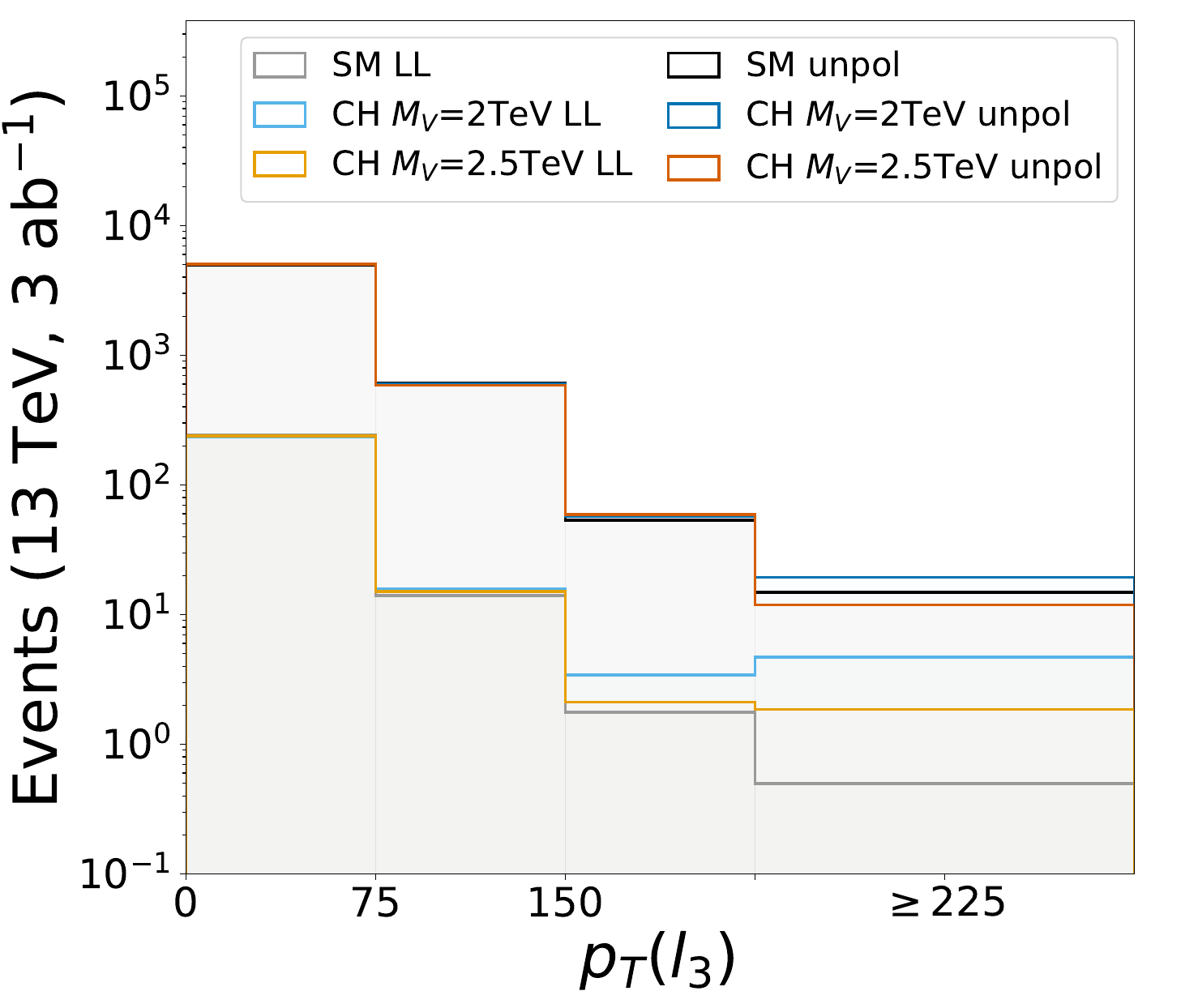}
    \caption{Distributions of the missing transverse energy and of the
    transverse momenta of the three leptons.
    Same samples, selection and normalisation as in Fig.~\ref{fig:masses}.}
    \label{fig:ptmet}
\end{figure}

\begin{figure}[t]
    \centering
        \includegraphics[width=0.45\linewidth]{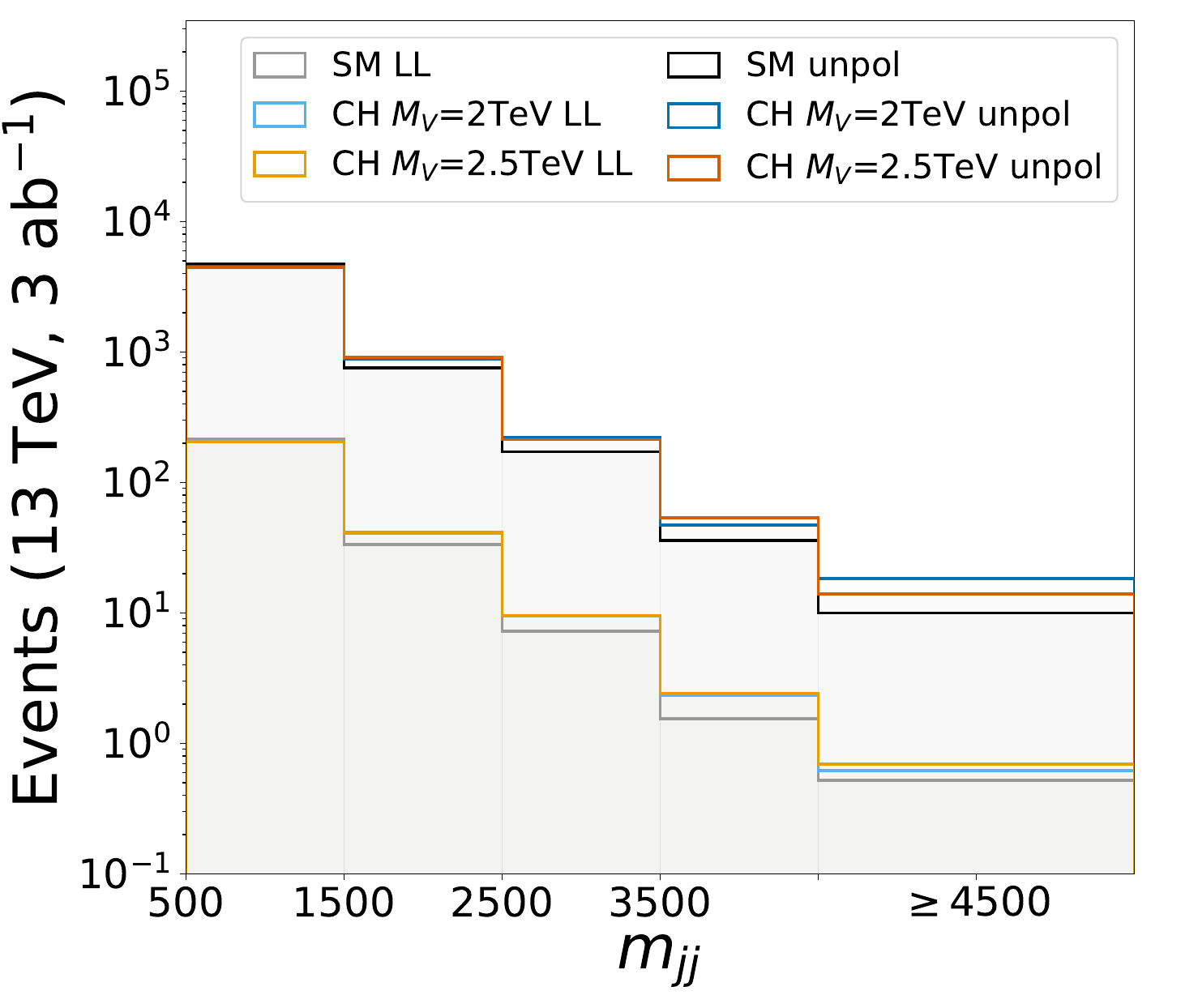}
        \includegraphics[width=0.45\linewidth]{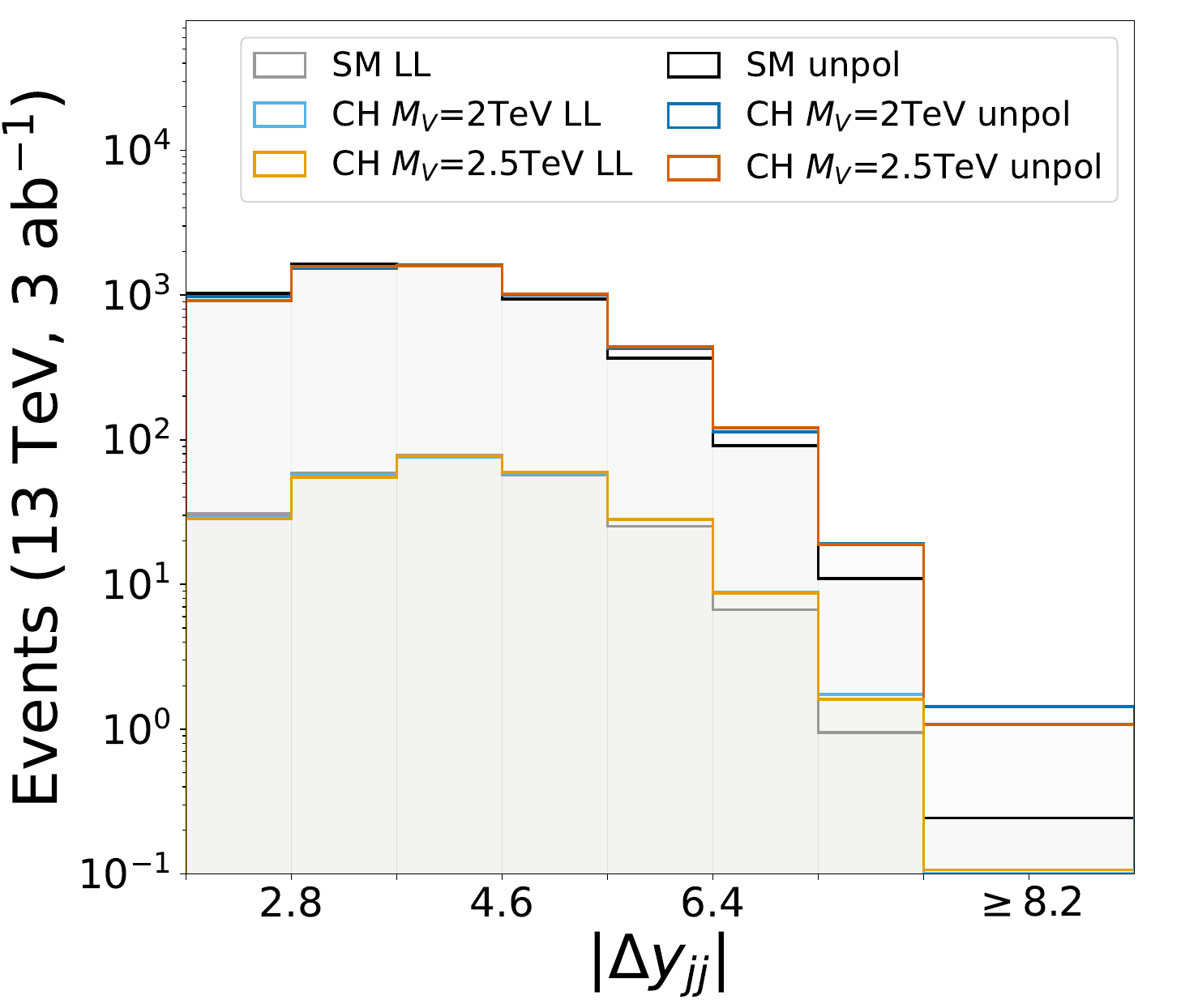}

        \includegraphics[width=0.45\linewidth]{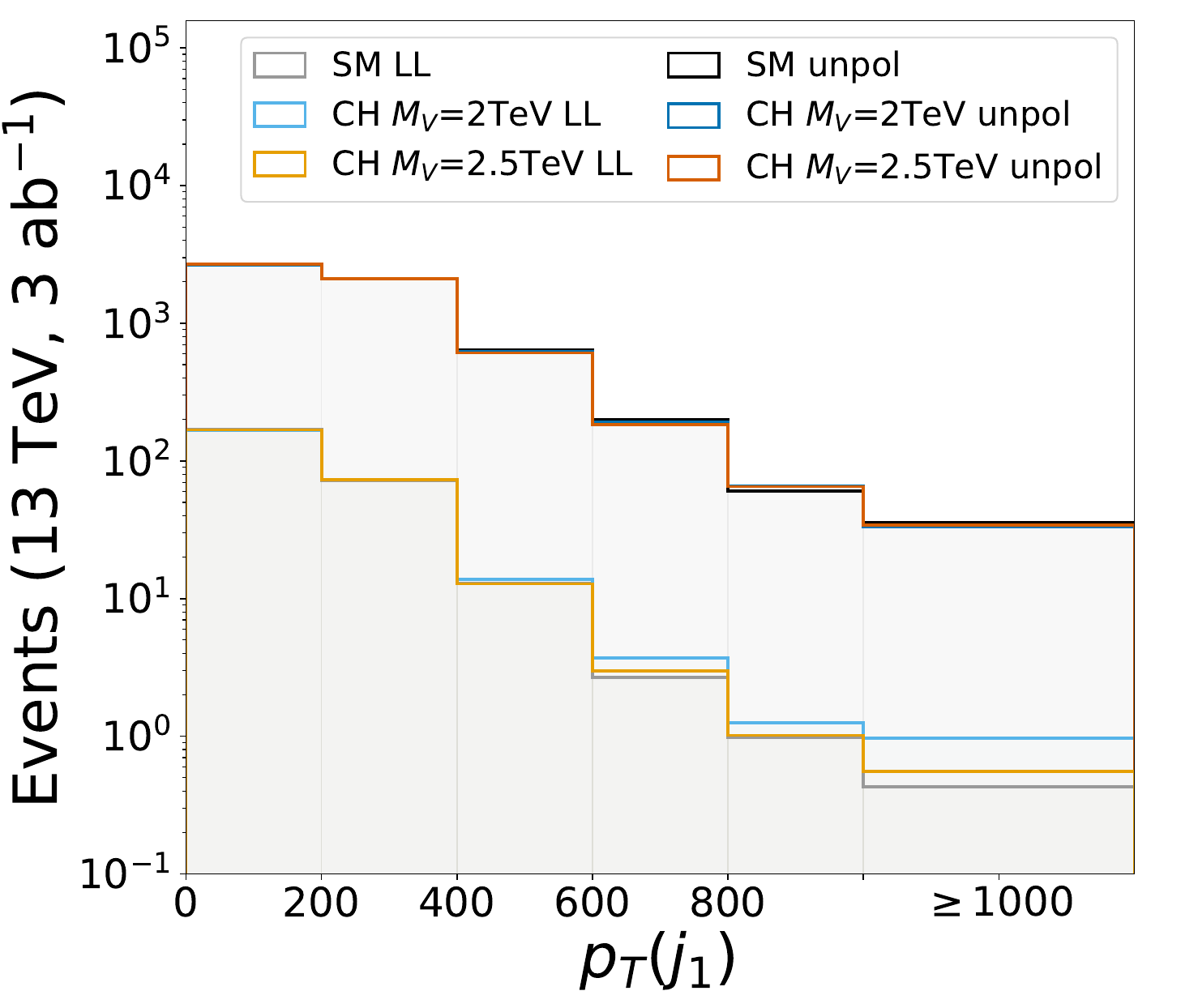}
        \includegraphics[width=0.45\linewidth]{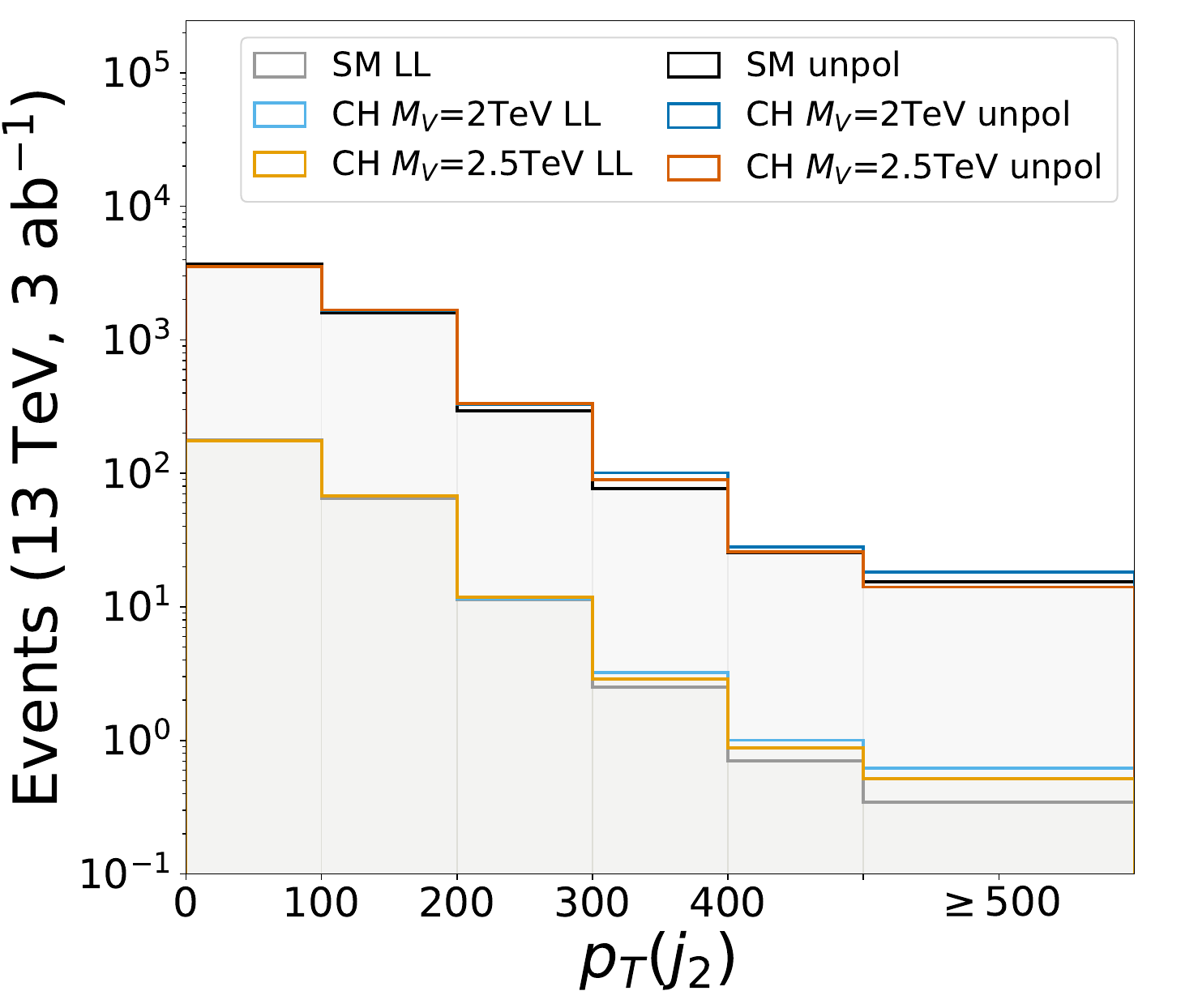}
    \caption{Distributions of the invariant mass and rapidity separation of
    the two tagging jets, and of their transverse momenta. Same samples,
    selection and normalization as in Fig.~\ref{fig:masses}. In these
    observables the CH and SM predictions differ only mildly.}
    \label{fig:jets}
\end{figure}

The results for the showered and reconstructed events are summarised in \cref{fig:masses,fig:ptmet,fig:jets}, which display the distributions of the lepton invariant masses, MET and lepton transverse momenta, and properties of the tagging jets, respectively.
The objects are ordered by $\pT$ such that e.g.\ $\ell_1$ is the hardest lepton.
The distributions show all events that passed the selection cuts in \cref{tab:cuts} normalised to 3~ab$^{-1}$.
The enhancement of the $W_L$ transverse momentum is reflected in the lepton invariant masses and momenta, see \cref{fig:masses,fig:ptmet}, which show significant deviations.
These are most pronounced in the tails of the distributions:
depending on the variable, the overflow bin is from about 3 times the Standard Model, in the case of $m_{\ell_1\ell_2}$ and $m_{\ell_1\ell_2\ell_3}$, to 9 times in $\pT(\ell_3)$. 
This reflects the modified high-energy behaviour of longitudinal $W_L Z_L$ scattering, sensitive to the exchange in the $s$-channel of the vector resonances.
For $M_V = 2$~TeV, deviations are already discernible in earlier parts of the tails of the distributions.
The larger vector mass decreases the contributions of the BSM propagators as these diagrams scale as $M_V^{-2}$. 
We therefore did not consider higher masses because the expected effect is even weaker.

For comparison, Fig.~\ref{fig:jets} shows properties of the two tagging jets: their invariant mass, rapidity difference and transverse momenta. 
In these observables, we do not see large differences between the CH model and the SM. This is not surprising, since the tagging jets come from the incoming quarks and are not directly affected by the pure VBS diagrams, which mainly concern the outgoing gauge bosons.

\section{Conclusion}
\label{sec:conc}

Vector boson scattering processes are very sensitive 
to the mechanism of electroweak symmetry breaking.
Longitudinally polarised vector bosons are of particular importance in this context, making them an interesting probe for composite Higgs models. 
We have focused on a class of models with an underlying fermionic description \cite{Ferretti:2016upr, Belyaev:2016ftv}. 
At tree level, the minimal cosets $\SU(4)/\Sp(4)$, $\SU(5)/\SO(5)$ and $\SU(4)^2/\SU(4)$ lead to identical results for the present study.
These models predict spin-1 resonances that mix with SM vector bosons and, thus, can be produced in $s$-channel processes at the LHC. 
Nevertheless, the corresponding bounds from existing Run-2 data still allow for scenarios in which the new states could have masses below 2~TeV. 

We presented a collider study of $W^{\pm}Zjj$ production at the LHC in the fully leptonic final state.
The composite Higgs model primarily modifies the couplings of the longitudinal polarisations of $W$ and $Z$.
We therefore compared kinematic distributions in the SM and the composite Higgs model for an unpolarised and a longitudinally polarised sample. 
The polarised case indeed showed significant deviations in the tails of the lepton invariant masses and transverse momenta, while the unpolarised case showed only small differences.
We studied benchmark points with a vector mass of 2~TeV, slightly above bounds from $s$-channel production, and 2.5~TeV to assess the mass dependence of the differences.
For the higher mass point, the CH distributions were still noticeably different from the SM, but the effect was less pronounced.
The properties of the tagging jets are mostly unchanged with respect to the SM. 
We also studied $W^+ W^+ jj$ production, which however did not show significant deviations.
The reason is that $W^{\pm}Zjj$ allows for an $s$-channel BSM resonance while $W^+ W^+ jj$ does not.

We have shown that composite Higgs models can noticeably change the kinematic distributions of the leptons in VBS processes.
These differences can be seen at the high-luminosity LHC but require an extraction of the longitudinally polarised events.
Promising variables are invariant masses and $\pT$ of the leptons. 
Although such deviations from the SM predictions prove the existence of new physics, they cannot pin down the underlying model or even the model class, as all models under study lead to the same effect in VBS processes.

It is well-known that higher order corrections are important in VBS processes, in particular for longitudinally polarised vector bosons, see e.g.~\cite{Biedermann:2017bss,Ballestrero:2018anz,Denner:2020zit,Denner:2021hsa,Dittmaier:2023nac,Denner:2024tlu,Denner:2024xul,Jager:2024sij,Denner:2025xdz,Denner:2026wst}. We expect the same to hold for composite Higgs models. QCD corrections should be straight forward but electroweak corrections will pose technical challenges due to the mixing of the spin-1 resonances with the SM vector bosons.

We restricted ourselves to the simplest case of only one scalar contributing to the EWSB, the Higgs boson.
However, all cosets feature an enlarged pNGB sector with additional neutral scalars that can in principle contribute to the breaking as well.
In that case, there will be more scalars contributing to the VBS processes, including $s$-channel resonances, as well as further spin-1 states.
For example, the $\SU(5)/\SO(5)$ coset features a doubly charged pNGB which then contributes to the $W^+W^+jj$ production.
We therefore expect larger deviations from the SM in this case.
We leave an investigation of such scenarios for future work.

\section*{Acknowledgements}
We thank Aryan Borkar, Jan Hadlik, Gia Khoriauli, and Raimund Str\"{o}hmer for useful discussions. 
We thank Moritz Volpert for his help in exploring the parameter space in the initial stages of this project, as well as Nico B\"{u}ttner for help with the event generation.

% TODO: include author contributions
%\paragraph{Author contributions}
%This is optional. If desired, contributions should be succinctly described in a single short paragraph, using author initials.

% TODO: include funding information
\paragraph{Funding information}
R.C.\ and W.P.\ have been supported by DFG project PO-1337/12-1. 
M.K.\ has been supported by DFG RTG 2994.

\appendix

\section{Mass matrices of the spin-1 states}
\label{app:model}

In this appendix we provide additional information on the mixing of the spin-1 resonances with the SM gauge bosons.
The following mass matrices coincide for all of the minimal cosets.
The Lagrangian for these models was constructed in ref.~\cite{Caliri:2024jdk} using the method of hidden local gauge symmetry. 
Its basic idea is to duplicate the $G\to H$ global symmetry breaking to $G_0 \times G_1 \to H_0 \times H_1$. 
In the $G_0$ sector only the generators corresponding to $\SU(2)_L \times \U(1)_Y$ are gauged with gauge coupling $\hat g$ and $\hat g'$, respectively. 
In the other sector $G_1$ is fully gauged (coupling $\tilde g$), giving rise to the spin-1 resonances.
Two symmetry breakings take place, $G_0 \to H_0$ and $G_1 \to H_1$, each with a set of pNGBs $\pi_{0,1}$ with decay constants $f_{0,1}$.
Only a linear combination of the pNGBs is physical, the remaining states are eaten to give mass to the spin-1 states in $G/H$. 
Finally, the remaining global symmetry is broken as $H_0 \times H_1 \to H$, yielding $\dim H$ would-be Goldstone bosons $K$ with decay constant $f_K$.
These give mass to the spin-1 states in $H$.
The pNGB decay constant $f_\pi$ of the physical pNGBs is given by
\begin{align}
   f_\pi = \sqrt{f^2_0 - r^2 f_1^2} = v_{\text{SM}}/\sin \theta \label{eq:pion-decay-constant}\, .
\end{align}
with an $\mathcal O(1)$ parameter $r$.
All decay constants are expected to be of the same order of magnitude.  
The misalignment angle $\sin\theta$ quantifies the ratio of the electroweak scale, given by the SM vacuum expectation value $v_\mathrm{SM} \simeq 246$~GeV.

From expanding the hidden symmetry Lagrangian \cite{Caliri:2024jdk}, one finds mass parameters of the vector states in $H$ and axial states in $G/H$ given by 
\begin{align}
    M_A^2 = \frac{f_1^2 \Tilde{g}^2}{2} \,, \qquad
    M_V^2 = \frac{f_K^2 \Tilde{g}^2}{2} \,.
\label{eq:mass-parameter}
\end{align}
These coincide with the masses of the states that do not participate in the mass mixing, while the mixing states have masses close to $M_V$ but with $\tilde g$-dependent corrections.
This follows from diagonalisation of the mass matrices, for the charged states in the basis $(\tilde W^+_\mu,\, a^+_\mu,\, v_{1\mu}^+,\, v_{2\mu}^+)$
\begin{align}
    \mathcal{M}_C^2 &= \mqty(
        \frac{ \hat g^2 M_{V}^2 (1 + \omega s^2_{\theta})}{ \tilde{g}^2} & \hat g \frac{r s_\theta M_A^2}{\sqrt{2}\Tilde{g}} & - \hat g \frac{M_V^2}{\sqrt{2} \Tilde{g}} & - \hat g \frac{M_V^2 c_\theta}{\sqrt{2} \Tilde{g}} \\
        \hat g \frac{r s_\theta M_A^2}{\sqrt{2}\Tilde{g}} & M_A^2 & 0&0 \\
        - \hat g \frac{M_V^2}{\sqrt{2} \Tilde{g}} & 0& M_V^2 & 0\\
        - \hat g \frac{M_V^2 c_\theta}{\sqrt{2} \Tilde{g}} &0 &0 & M_V^2
    ) \,
 \label{eq:MC}   
\end{align}
where $\omega=(f_0^2 / f_K^2 - 1)/2$ and we abbreviate $s_\theta = \sin\theta$ and $c_\theta = \cos\theta$.
In the neutral sector we take the basis $(B_\mu, \tilde W^3_\mu,\, a^0_\mu,\, v_{1\mu}^0,\, v_{2\mu}^0)$, which yields the mass matrix
\begin{align}
    \mathcal{M}_N^2 &= \mqty(
        \frac{\hat g'^2 M_{V}^2 (1 + \omega s^2_{\theta})}{ \tilde{g}^2} & -\frac{\hat g' \hat g M_{V}^2 \omega  s^2_{\theta}}{\tilde{g}^2} & - \hat g' \frac{r s_\theta M_A^2}{\sqrt{2} \Tilde{g}} & - \hat g' \frac{M_V^2}{\sqrt{2} \Tilde{g}} &  \hat g' \frac{M_V^2 c_\theta}{\sqrt{2} \Tilde{g}} \\
        -\frac{ \hat g'  \hat g M_{V}^2 \omega  s^2_{\theta}}{\tilde{g}^2} & \frac{ \hat g^2 M_{V}^2 (1 + \omega s^2_{\theta})}{ \tilde{g}^2} &  \hat g \frac{r M_A^2 s_\theta}{\sqrt{2} \Tilde{g}} & - \hat g \frac{M_V^2}{\sqrt{2} \Tilde{g}} & - \hat g \frac{M_V^2 c_\theta}{\sqrt{2} \Tilde{g}} \\
        - \hat g' \frac{r s_\theta M_A^2}{\sqrt{2} \Tilde{g}} &  \hat g \frac{r M_A^2 s_\theta}{\sqrt{2} \Tilde{g}} & M_A^2 & 0&0 \\
        - \hat g' \frac{M_V^2}{\sqrt{2} \Tilde{g}} & - \hat g \frac{M_V^2}{\sqrt{2} \Tilde{g}} &0 & M_V^2 & 0\\
         \hat g' \frac{M_V^2 c_\theta}{\sqrt{2} \Tilde{g}} & - \hat g \frac{M_V^2 c_\theta}{\sqrt{2} \Tilde{g}} & 0&0 & M_V^2 ) \,.
\label{eq:MN}         
\end{align}
In this sector, the two states $v_{1\mu}^0$ and $v_{2\mu}^0$ mix with the photon and the $Z$-boson, whereas the mixing with $a_{\mu}^0$ is suppressed and vanishes in the limit $\sin\theta\to 0$.
We denote the corresponding heavy mass eigenstates that sizeably mix with the electroweak vector bosons by $V^0_{1\mu}$ and $V^0_{2\mu}$.

\section{Results for $W^+ W^+jj$ production}
\label{app:WW}

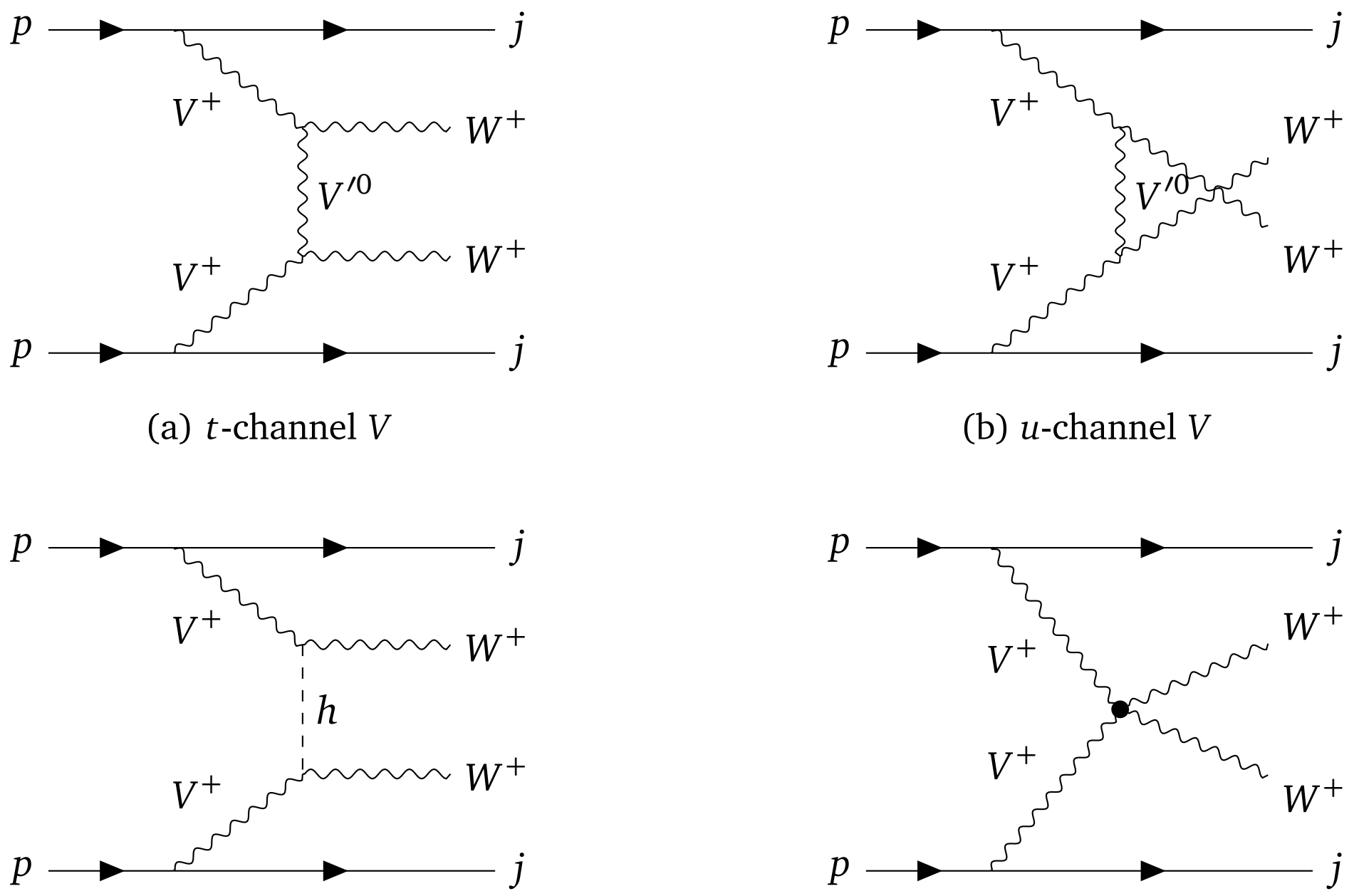
\begin{figure}[t]
  \centering
  \begin{subfigure}{0.45\textwidth}
    \centering
    \begin{tikzpicture}
      \begin{feynman}
        \vertex (i1) at (0, 3) {\(p\)};
        \vertex (v1) at (1.4, 3);
        \vertex (f1) at (4.6, 3) {\(j\)};
        \vertex (i2) at (0, 0) {\(p\)};
        \vertex (v2) at (1.4, 0);
        \vertex (f2) at (4.6, 0) {\(j\)};
        \vertex (t1) at (2.6, 2.1);
        \vertex (t2) at (2.6, 0.9);
        \vertex (w1) at (4.4, 2.1) {\(W^+\)};
        \vertex (w2) at (4.4, 0.9) {\(W^+\)};
        \diagram* {
          (i1) -- [fermion] (v1) -- [fermion] (f1),
          (i2) -- [fermion] (v2) -- [fermion] (f2),
          (v1) -- [boson, edge label'=\(V^+\)] (t1),
          (v2) -- [boson, edge label=\(V^+\)] (t2),
          (t1) -- [boson, edge label=\(V'^0\)] (t2),
          (t1) -- [boson] (w1),
          (t2) -- [boson] (w2),
        };
      \end{feynman}
    \end{tikzpicture}
    \caption{$t$-channel $V$}
    \label{fig:ww_t}
  \end{subfigure}
  \qquad
  \begin{subfigure}{0.45\textwidth}
    \centering
    \begin{tikzpicture}
      \begin{feynman}
        \vertex (i1) at (0, 3) {\(p\)};
        \vertex (v1) at (1.4, 3);
        \vertex (f1) at (4.6, 3) {\(j\)};
        \vertex (i2) at (0, 0) {\(p\)};
        \vertex (v2) at (1.4, 0);
        \vertex (f2) at (4.6, 0) {\(j\)};
        \vertex (t1) at (2.6, 2.1);
        \vertex (t2) at (2.6, 0.9);
        \vertex (w1) at (4.4, 2.1) {\(W^+\)};
        \vertex (w2) at (4.4, 0.9) {\(W^+\)};
        \diagram* {
          (i1) -- [fermion] (v1) -- [fermion] (f1),
          (i2) -- [fermion] (v2) -- [fermion] (f2),
          (v1) -- [boson, edge label'=\(V^+\)] (t1),
          (v2) -- [boson, edge label=\(V^+\)] (t2),
          (t1) -- [boson, edge label=\(V'^0\)] (t2),
          (t1) -- [boson] (w2),
          (t2) -- [boson] (w1),
        };
      \end{feynman}
    \end{tikzpicture}
    \caption{$u$-channel $V$}
    \label{fig:ww_u}
  \end{subfigure}
  \\[1.5em]
  \begin{subfigure}{0.45\textwidth}
    \centering
    \begin{tikzpicture}
      \begin{feynman}
        \vertex (i1) at (0, 3) {\(p\)};
        \vertex (v1) at (1.4, 3);
        \vertex (f1) at (4.6, 3) {\(j\)};
        \vertex (i2) at (0, 0) {\(p\)};
        \vertex (v2) at (1.4, 0);
        \vertex (f2) at (4.6, 0) {\(j\)};
        \vertex (t1) at (2.6, 2.1);
        \vertex (t2) at (2.6, 0.9);
        \vertex (w1) at (4.4, 2.1) {\(W^+\)};
        \vertex (w2) at (4.4, 0.9) {\(W^+\)};
        \diagram* {
          (i1) -- [fermion] (v1) -- [fermion] (f1),
          (i2) -- [fermion] (v2) -- [fermion] (f2),
          (v1) -- [boson, edge label'=\(V^+\)] (t1),
          (v2) -- [boson, edge label=\(V^+\)] (t2),
          (t1) -- [scalar, edge label=\(h\)] (t2),
          (t1) -- [boson] (w1),
          (t2) -- [boson] (w2),
        };
      \end{feynman}
    \end{tikzpicture}
    \caption{$t$-channel Higgs}
    \label{fig:ww_h}
  \end{subfigure}
  \qquad
  \begin{subfigure}{0.45\textwidth}
    \centering
    \begin{tikzpicture}
      \begin{feynman}
        \vertex (i1) at (0, 3) {\(p\)};
        \vertex (v1) at (1.4, 3);
        \vertex (f1) at (4.6, 3) {\(j\)};
        \vertex (i2) at (0, 0) {\(p\)};
        \vertex (v2) at (1.4, 0);
        \vertex (f2) at (4.6, 0) {\(j\)};
        \vertex [dot] (q) at (2.6, 1.5) {};
        \vertex (w1) at (4.4, 2.3) {\(W^+\)};
        \vertex (w2) at (4.4, 0.7) {\(W^+\)};
        \diagram* {
          (i1) -- [fermion] (v1) -- [fermion] (f1),
          (i2) -- [fermion] (v2) -- [fermion] (f2),
          (v1) -- [boson, edge label'=\(V^+\)] (q),
          (v2) -- [boson, edge label=\(V^+\)] (q),
          (q) -- [boson] (w1),
          (q) -- [boson] (w2),
        };
      \end{feynman}
    \end{tikzpicture}
    \caption{quartic gauge coupling}
    \label{fig:ww_q}
  \end{subfigure}
  \caption{Feynman diagrams of the vector boson scattering processes for
    $p p \to W^+W^+ j j$. $V^+$ and $V'^0$ can be either a SM vector boson or a
    spin-1 resonance. Since the $W^+W^+$ system carries electric charge two,
    there is no $s$-channel contribution.}
  \label{fig:ww_diagrams}
\end{figure}

We present here the results  for $W^+ W^+jj$ production for the first benchmark point which is
characterized by $M_V = 2$~TeV, $g_{V\pi\pi} = \tilde g = 4$ and
$f = 1$~TeV. In case of the second benchmark point with  $M_V = 2.5$~TeV the effects are even less pronounced. In the following both $W$ bosons decay leptonically, $W^+ \to \ell^+\nu$ ($\ell = e, \mu, \tau$).
The corresponding Feynman diagrams are shown in \cref{fig:ww_diagrams}.

For the implementation of the analysis, we follow \cite{ATLAS:2023sua}. 
The object definition for this analysis is the same as in \cref{tab:objects}. In \cref{tab:ww_selection} instead we summarise the event selection for  $pp \to W^+W^+jj$. 
Here we require two signal leptons with the same electric charge and veto events with an additional baseline lepton forming a
$Z$  with a signal lepton, $|m_{\ell\ell} - m_Z| < 15$~GeV. 
We also veto $Z \to ee$ candidates among the two signal electrons, and require $m_{\ell\ell} > 20$~GeV. The requirements on the jets, on
$E_T^{\text{miss}}$, the $b$-jet veto and the VBS cuts $m_{jj} > 500$~GeV and
$|\Delta y_{jj}| > 2$ are the same as for $W^\pm Z jj$.

\begin{table}
  \centering
  \begin{tabular}{rl}
    \toprule
    & $W^+W^+jj$\\
    \midrule
Leptons
      & $N_\ell = 2$, same electric charge \\
    \multirow{2}{*}{Third-lepton veto}
      & no additional SF baseline lepton with \\
      & $|m_{\ell\ell} - \mZ| < 15\GeV$ with a signal lepton \\
    $Z \to ee$ veto
      & $|m_{ee} - \mZ| > 15\GeV$ if $\ell_1, \ell_2 = e$ \\
    Dilepton mass & $m_{\ell\ell} > 20\GeV$ \\
  Jets
      & $N_j \geq 2$, $\pT(j_1) > 65\GeV$, $\pT(j_2) > 35\GeV$ \\
    $MET$ & $> 30\GeV$ \\
    $b$-jet veto & $N_b = 0$ \\
    VBS topology
      & $m_{j_1 j_2} > 500\GeV$, $|\Delta y_{j_1 j_2}| > 2$ \\
      \midrule
  \end{tabular}
  \caption{Summary of the event selection for
  $pp \to W^+W^+jj$, adapted from Ref.~\cite{ATLAS:2023sua}. SF stands for same flavour. }
  \label{tab:ww_selection}
\end{table}

\begin{figure}[htbp]
    \centering
        \includegraphics[width=0.45\linewidth]{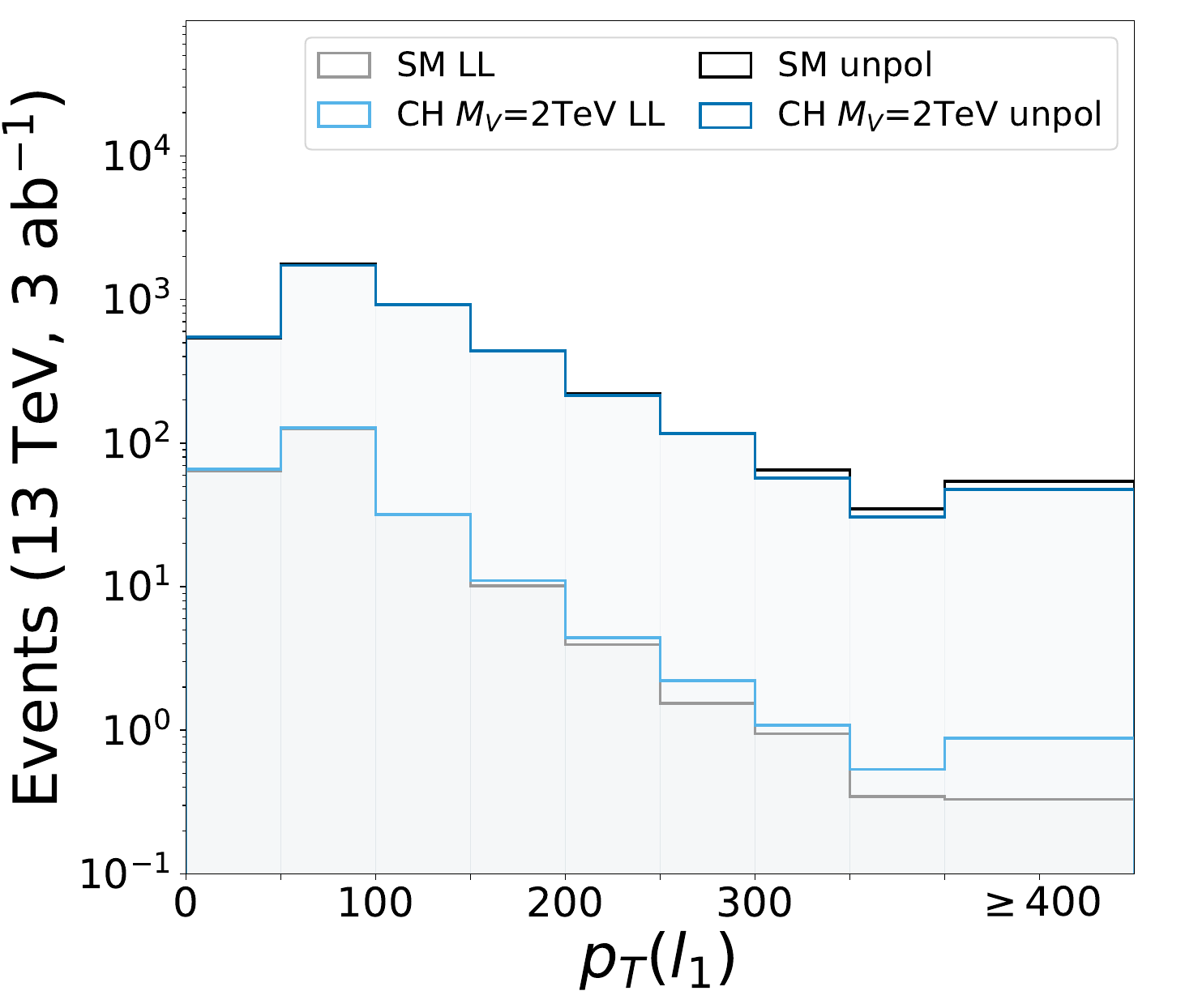}
        \includegraphics[width=0.45\linewidth]{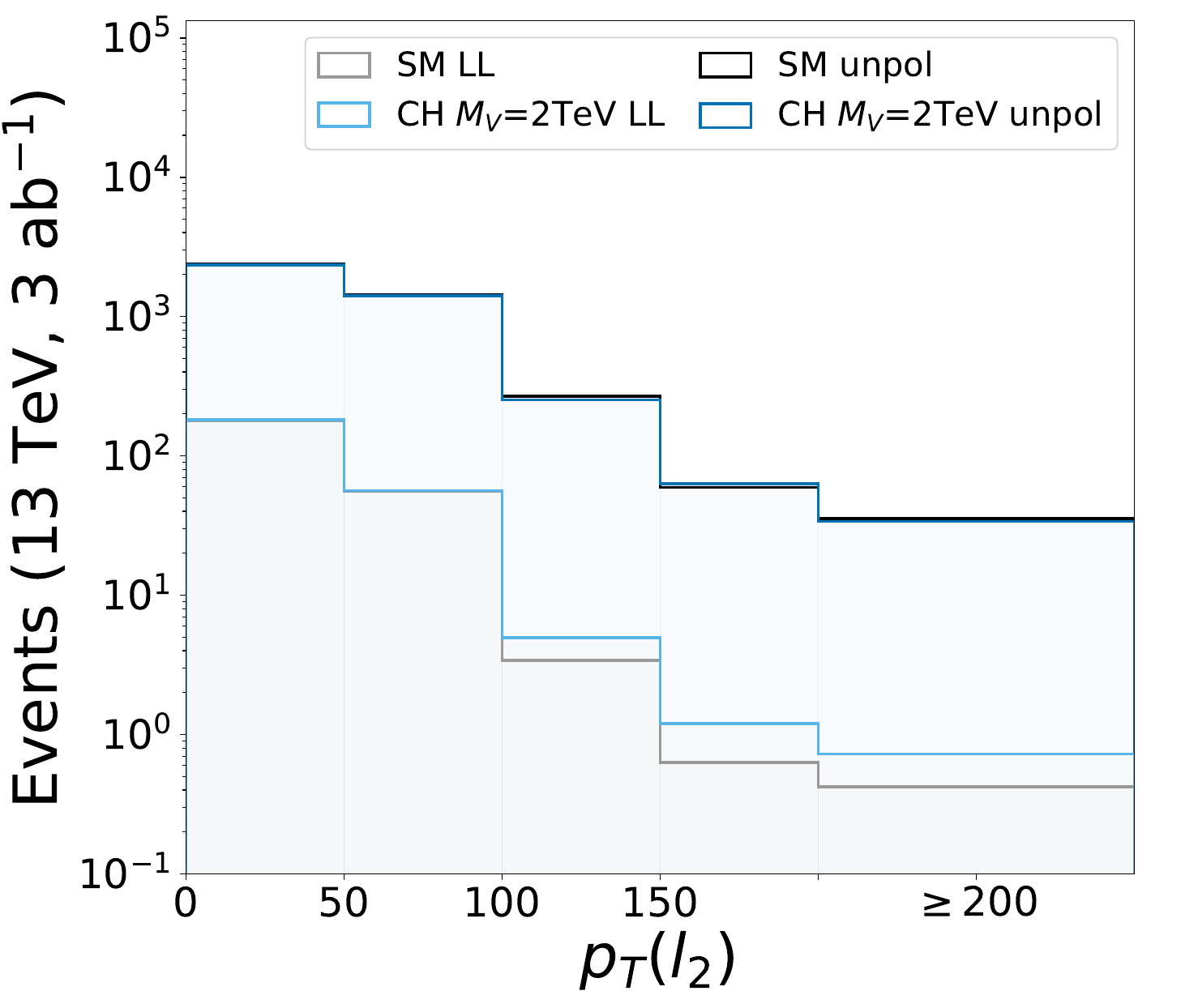}

        \includegraphics[width=0.45\linewidth]{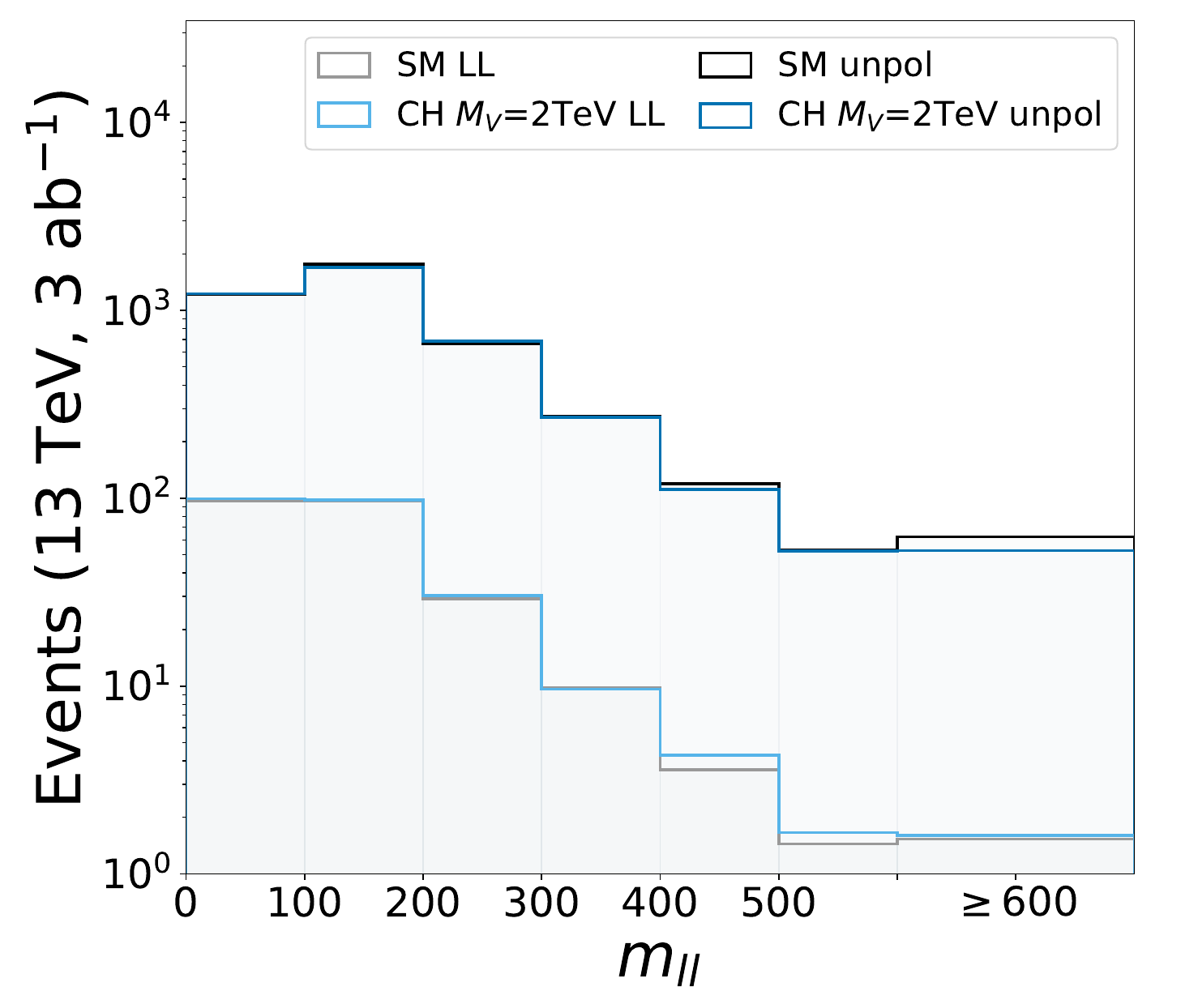}
        \includegraphics[width=0.45\linewidth]{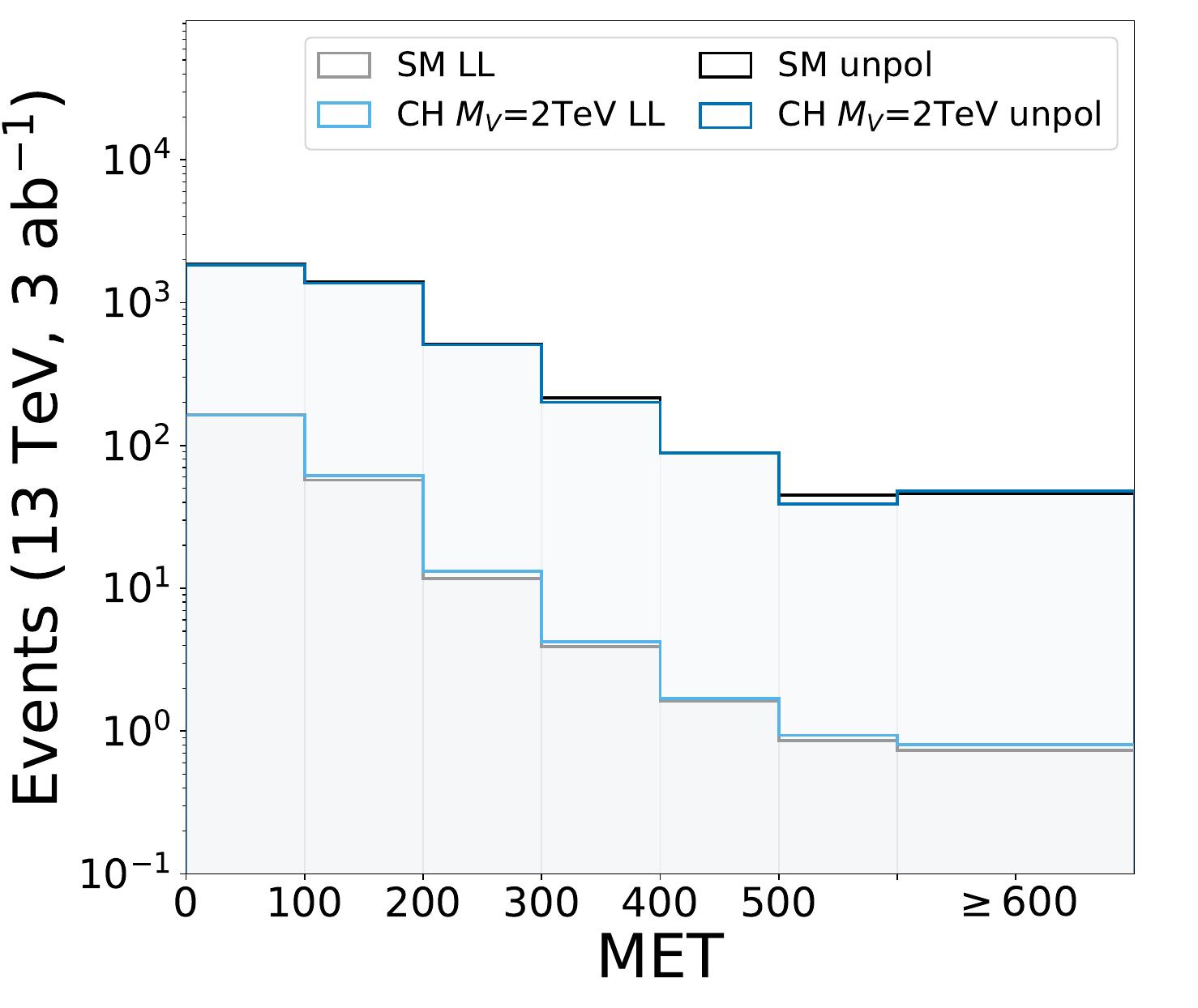}
    \caption{Distributions of the lepton transverse momenta, the dilepton
    invariant mass and the missing transverse energy in
    $pp \to W^+W^+jj \to \ell^+\ell^{\prime +}\,\nu \nu'\,jj$ after the full event selection.
    The black/grey lines correspond to the SM and the blue ones to CH with
    $M_V = 2$~TeV, $g_{V\pi\pi} = \tilde g = 4$ and $f = 1$~TeV. Dark lines show the
    unpolarised samples, and light lines the LL samples. Each curve is
    normalised to the expected SM yield at
    $\sqrt{s} = 13$~TeV and $3~\text{ab}^{-1}$. The last bin is an overflow
    bin.}
    \label{fig:ww}
\end{figure}

In \cref{fig:ww} we show the histograms obtained for this channel. The distributions are
normalised as before to the expected SM number of events at $\sqrt{s} = 13$~TeV and
$3~\text{ab}^{-1}$ with $\sigma_{\text{SM}} = 27.5$~fb (1.75~fb) for the unpolarised (LL) sample. 
The effects are small in this case compared to the $W^\pm Z jj$ channel.
In particular, there are hardly any differences for unpolarised vector bosons. 
In case of longitudinally polarised vector bosons we observe a small enhancement just in the tails of the $\pT$ distributions of the leptons. 
We have checked that the other observables don't give noticeable differences compared to SM expectations. 
The reason is a cancellation between the contributions of the $t$- and $u$-channels diagrams and the quartic contributions, see \cref{fig:ww_diagrams}.

\clearpage


% Provide your bibliography here. You have two options:

%%% FIRST OPTION
% Write your entries here directly, following the example below, including:
% Author(s), Title, Journal Ref. with year in parentheses at the end, followed by the DOI number.

%\begin{thebibliography}{99}
%\bibitem{1931_Bethe_ZP_71} H. A. Bethe, {\it Zur Theorie der Metalle. i. Eigenwerte und Eigenfunktionen der linearen Atomkette}, Zeit. f{\"u}r Phys. {\bf 71}, 205 (1931), \doi{10.1007\%2FBF01341708}.
%\bibitem{arXiv:1108.2700} P. Ginsparg, {\it It was twenty years ago today... }, \url{http://arxiv.org/abs/1108.2700}.
%\end{thebibliography}

% Use your bibtex library, formatted by the SciPost style file.
\bibliography{literature.bib}

%%%%%%%%%% END TODO: BIBLIOGRAPHY

\end{document}